\documentclass[a4paper,11pt]{article}
\pdfoutput=1

\usepackage{jheppub}
\usepackage[T1]{fontenc} 
\usepackage{amsmath,amssymb,amsbsy,amstext,amsthm,simplewick,amsfonts,tensor} 
\usepackage{graphicx}

\def\l{\left}
\def\r{\right}
\def\({\l(}
\def\){\r)}
\def\[{\l[}
\def\]{\r]}
\def\pd{\partial}
\def\t{\tensor}

\def\til{\tilde}
\def\cal{\mathcal}

\def\sep{~,\quad}
\def\la{\langle}
\def\ra{\rangle}

\def\j{\varphi}
\def\f{\phi}

\DeclareMathOperator{\sinc}{sinc}

\def\diag{\mathrm{diag}}

\def\d{\mathrm d} 

\title{\boldmath Gravitational effects on oscillon lifetimes}

\author{Hong-Yi Zhang}
\affiliation{Department of Physics and Astronomy, Rice University, Houston, TX 77005, USA}

\emailAdd{hongyi@rice.edu}

\abstract{Many scalar field theories with attractive self-interactions support exceptionally long-lived, spatially localized and time-periodic field configurations called oscillons. A detailed study of their longevity is important for understanding their applications in cosmology. In this paper, we study gravitational effects on the decay rate and lifetime of dense oscillons, where self-interactions are more or at least equally important compared with gravitational interactions. As examples, we consider the $\alpha$-attractor T-model of inflation and the axion monodromy model, where the potentials become flatter than quadratic at large field values beyond some characteristic field distance $F$ from the minimum. For oscillons with field amplitudes of $\mathcal{O}(F)$ and for $F\ll 0.1 M_\mathrm{pl}$, we find that their evolution is almost identical to cases where gravity is ignored. For $F\sim 0.1 M_\mathrm{pl}$, however, including gravitational interactions reduces the lifetime slightly.}

\begin{document}
\maketitle
\flushbottom

\section{Introduction}
Oscillons are exceptionally long-lived, spatially-localized and time-periodic field configurations that exist in real-valued scalar field theories with attractive self-interactions \cite{osti_4051808,Bogolyubsky:1976yu,Copeland:1995fq,Amin:2010jq,Amin:2013ika}. The natural emergence of oscillons from general initial conditions \cite{Seidel:1993zk, Farhi:2007wj, Amin:2010xe, Levkov:2018kau} makes them relevant to various cosmological contexts, e.g. reheating after inflation \cite{Amin:2010dc,Amin:2011hj, Lozanov:2014zfa, Lozanov:2016hid, Lozanov:2017hjm,Kou:2019bbc}, scalar dark matter \cite{Olle:2019kbo,Amin:2019ums, Arvanitaki:2019rax, Kawasaki:2020jnw}, phase transitions in the early universe \cite{Dymnikova:2000dy, Gleiser:2010qt, Bond:2015zfa}, gravitational wave production \cite{Zhou:2013tsa, Helfer:2018vtq, Liu:2017hua, Amin:2018xfe,Lozanov:2019ylm, Dietrich:2018jov}, electromagnetic bursts \cite{Dietrich:2018jov, Hook:2018iia, Clough:2018exo, Levkov:2020txo, Prabhu:2020yif, Amin:2020vja}, the 21cm forest \cite{Kawasaki:2020tbo}, and even formation of black holes \cite{Helfer:2016ljl, Muia:2019coe, Nazari:2020fmk, Widdicombe:2019woy} and primordial black holes \cite{Khlopov:1985jw, Cotner:2018vug, Cotner:2019ykd}. For a quantitative understanding of their relevance to cosmology, a detailed study of oscillon decay rates and lifetimes is important.

\begin{figure}
	\centering
	\begin{minipage}{0.485\linewidth}
		\includegraphics[width=\linewidth]{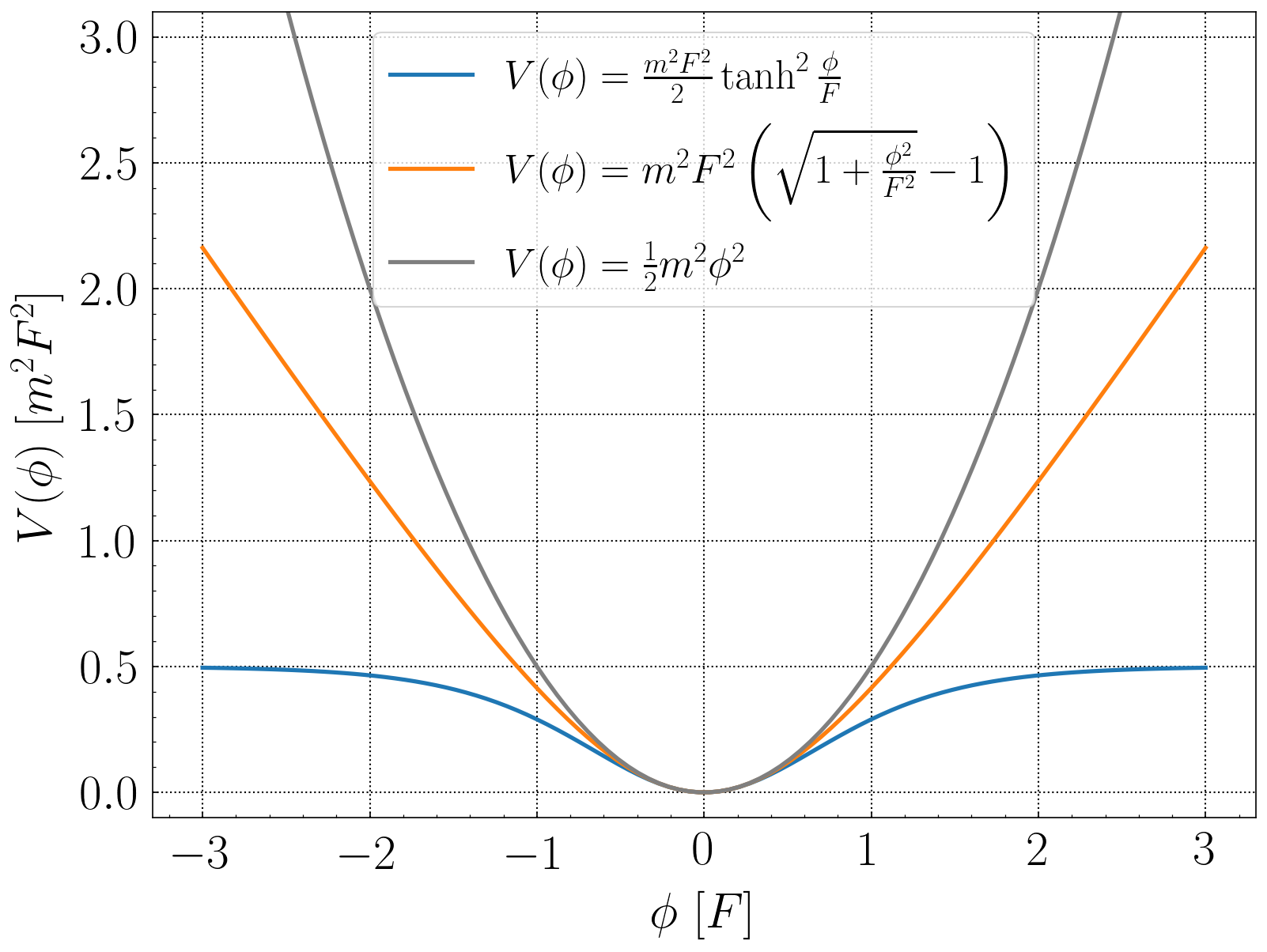}
	\end{minipage}
	\begin{minipage}{0.495\linewidth}
		\includegraphics[width=\linewidth]{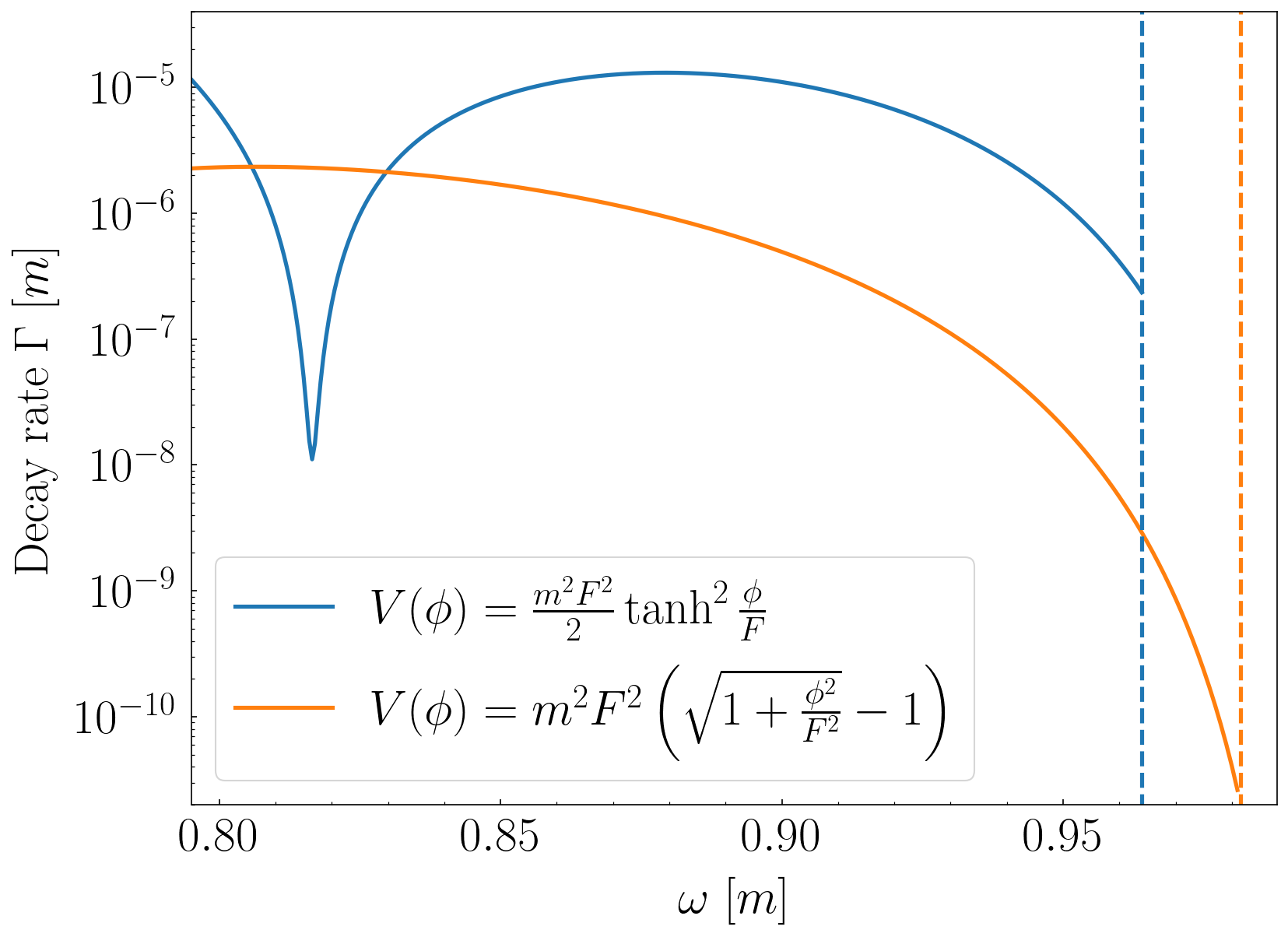}
	\end{minipage}
	\caption{\small We will consider oscillons in two flattened potentials in this paper (left panel), and semi-analytical calculation of their decay rate is presented for Minkowski spacetime (right panel). The dashed lines indicate the location of $\omega_\mathrm{crit}$, at which oscillons finally decay away due to instabilities to small fluctuations.}
	\label{fig:decayrateminkowski}
\end{figure}

In the absence of gravity, the decay rate of small-amplitude oscillons \cite{Segur:1987mg,Fodor:2008du,Hertzberg:2010yz} and large-amplitude ones in some polynomial models \cite{Mukaida:2016hwd,Ibe:2019vyo} has been characterized in detail. In a recent paper \cite{Zhang:2020bec}, we calculated the decay rate of oscillons without restricting them to small amplitudes or to polynomial potentials. By investigating the relation between decay rates and fundamental frequencies of oscillons in a given theory, see the right panel of figure \ref{fig:decayrateminkowski}, we found that two major features were responsible for their particularly long lifetimes: (1) Some exceptionally stable intermediate configurations exist, e.g. a dip structure of the blue curve around $\omega_\mathrm{dip}\approx 0.82 m$. For such configurations, the leading decay channel into scalar radiation vanishes. (2) The decay rate is dramatically suppressed just before their final collapse at the critical frequency $\omega_\mathrm{crit}\approx 0.98 m$, such as the orange curve. 

The gravitational effects on oscillon decay rates are expected to depend on the relative magnitude between gravitational and self- interactions. In 3+1 dimensions, the lifetime of small-amplitude \emph{dilute} oscillons (whose self-interactions are negligible) was shown to exceed the present age of the universe \cite{Grandclement:2011wz,Eby:2015hyx, Eby:2020ply}.\footnote{The stability of dilute oscillons is ensured by gravitational attraction, for example, oscillatons \cite{Seidel:1991zh,UrenaLopez:2002gx,Alcubierre:2003sx} and dilute axion stars \cite{Visinelli:2017ooc, Eby:2019ntd}. Their size can be cosmological scales, e.g. $\sim \mathrm{kpc}$ for fuzzy dark matter \cite{Hui:2016ltb}.} What was not clear to us is whether the existence of gravity stablizes large-amplitude \emph{dense} ones, whose self-interactions are more or at least equally important. In this paper, we will study two well-motivated examples to explore this impact, the $\alpha$-attractor T-model of inflation \cite{Kallosh:2013hoa,Lozanov:2017hjm} and the axion monodromy model \cite{Silverstein:2008sg,Amin:2011hj,McAllister:2014mpa} 
\begin{align}\label{potentials}
V(\f) = \frac{m^2F^2}{2}\tanh^2 \frac{\f}{F} \sep V(\f) = m^2F^2\( \sqrt{1+ \frac{\phi^2}{F^2} } - 1 \) ~,
\end{align}
where $F$ is the amplitude scale that indicates a significant deviation from a quadratic minimum, see the left panel of figure \ref{fig:decayrateminkowski}. 

Throughout the paper we will assume classical field theory and spherical symmetry. The first assumption is a standard procedure to deal with compact objects like dense oscillons due to their particularly large occupation number \cite{Guth:2014hsa}. We also stick to the second one because it simplifies the problem quite a bit and non-spherical components usually decay rapidly with little disturbance on oscillon evolution \cite{Seidel:1993zk,Hindmarsh:2006ur}. Even if the early inhomogeneities forming oscilllons possess angular momentum,  a non-axisymmetric instability seems to develop and ejects all the angular momentum from the scalar star \cite{Sanchis-Gual:2019ljs}.

The rest of this paper is organized as follows. In section \ref{sec:Minkowski}, we briefly review the method developed in \cite{Zhang:2020bec} to calculate oscillon decay rates and lifetimes in Minkowski spacetime. In section \ref{sec:GR}, we introduce a Lagrangian mechanism and define the mass and energy of oscillons in curved spacetime. In section \ref{sec:linearized_gravity}, we generalize the method to calculate oscillon decay rates and lifetimes in the non-relativistic limit and weak-field limit of gravity. Then we discuss the gravitational effects on oscillon lifetimes in section \ref{sec:examples} and make conclusions in section \ref{sec:summary}. In appendix \ref{sec:potential_expansion} and \ref{sec:GR_numerics}, we provide analytical expressions for cosine series of the scalar potential and describe numerical algorithms for full GR simulations respectively. We will adopt the natural units $c=\hbar=1$, and frequently use the reduced Planck mass defined by $M_\mathrm{pl}\equiv (8\pi G)^{-1/2}\sim 10^{18} \mathrm{GeV}$.

\section{Oscillons in Minkowski spacetime}
\label{sec:Minkowski}
We study oscillon dynamics in Minkowski spacetime in this section. In section \ref{sec:Minkowski_radiation}, we briefly review a model-independent calculation of oscillon decay rates and lifetimes following our earlier work \cite{Zhang:2020bec}. Then in section \ref{sec:Minkowski_virial}, we derive a virial theorem and introduce some small parameters that will allow us to simplify equations when gravitational effects are included.

\subsection{Profiles, decay rates and lifetimes}
\label{sec:Minkowski_radiation}
Let us begin with a real-valued scalar field with the Lagrangian given by
\begin{align}
\cal L = -\frac{1}{2} g^{\mu\nu} \pd_\mu\f \pd_\nu\f - \frac{1}{2}m^2\f^2 - V_\mathrm{nl}(\f) ~,
\end{align}
where $g_{\mu\nu}=\diag(-1,1,1,1)$ is the Minkowski metric and $V_\mathrm{nl}(\f)$ is the nonlinear part of the potential $V(\f)$. The equation of motion is the Klein-Gordon equation
\begin{align}\label{EOM}
\[ -\pd_t^2 + \nabla^2 - m^2 \] \f - V_\mathrm{nl}'(\f) = 0 ~,
\end{align}
where $\nabla^2\equiv \pd_r^2+(2/r)\pd_r$. For simplicity we will only consider symmetric potentials, but the method developed in this paper should also be applicable to asymmetric ones. As suggested in \cite{Seidel:1991zh}, we approximate oscillons by a cosine series
\begin{align}\label{profile_single_freq}
\f(t,r) = \f_\mathrm{osc}(t,r) + \xi(t,r) = \phi_1(r)\cos(\omega t) + \sum_{j=3}^{\infty}\xi_j(r) \cos(j\omega t) ~,
\end{align}
where $j$ is odd, $\f_\mathrm{osc}$ is a single-frequency profile and $\xi(t,r)$ includes all the radiating modes.\footnote{The $\xi_j(r)\cos(j\omega t)$ mode is a radiating mode if $j\omega>m$. Notice that this expansion is actually a balance between ingoing and outgoing waves, and we must manually ignore the ingoing contributions in the end.} Typically $|\xi|\ll|\f_\mathrm{osc}|$ inside oscillons.\footnote{\label{footnote:single_freq_profile}The assumption of a single-frequency profile becomes invalid when $\omega\ll m$, however we never consider the $\omega\ll m$ limit because: (1) The particles that make up the oscillons in this regime are  relativistic as you will see in section \ref{sec:Minkowski_virial}. In this case, a large number of particles can easily pop in and out of the condensate and we are unlikely to have a stable long-lived condensate. (2) The radiating modes, and hence decay rates, are typically too large for the oscillon to maintain a stable configuration and (3) The size of the oscillon approaches the Schwarzschild radius and the nonlinearity of gravity becomes important, which is beyond the scope of this paper.} As a result, the potential and its derivatives can also be written in terms of the Fourier cosine series
\begin{align}\label{V_expansion}
U&\equiv V_\mathrm{nl}(\f_\mathrm{osc}) = \frac{1}{2}U_0(r) + \sum_{j=2}^{\infty} U_j(r)\cos(j\omega t) \quad\text{and}\quad U_j = \frac{\omega}{\pi} \int_{-\frac{\pi}{\omega}}^{\frac{\pi}{\omega}} V_\mathrm{nl}(\f_\mathrm{osc}) \cos(j\omega t) dt ~,\\
\label{M_expansion}
M &\equiv V_\mathrm{nl}''(\f_\mathrm{osc}) = \frac{1}{2}M_0(r) + \sum_{j=2}^{\infty} M_j(r)\cos(j\omega t) \quad\text{and}\quad M_j = \frac{\omega}{\pi} \int_{\frac{\pi}{\omega}}^{\frac{\pi}{\omega}} V_\mathrm{nl}''(\f_\mathrm{osc}) \cos(j\omega t) dt ~.
\end{align}
where $j$ is even, and 
\begin{align}
\label{J_expansion}
J\equiv V_\mathrm{nl}'(\f_\mathrm{osc}) = \sum_{j=1}^{\infty} J_j(r)\cos(j\omega t) &\quad\text{and}\quad J_j = \frac{\omega}{\pi} \int_{-\frac{\pi}{\omega}}^{\frac{\pi}{\omega}} V_\mathrm{nl}'(\f_\mathrm{osc}) \cos(j\omega t) dt ~,
\end{align}
where $j$ is odd. The $U_j$ is a functional of $\f_1$ hence a function of $r$, namely $U_j(\f_1)\equiv U_j(r)$. We will mix the notation $U_j(\f_1)$ and $U_j(r)$, and similarly for $J_j$ and $M_j$. For polynomial potentials, it is possible to find analytical expressions for $U_j,M_j$ and $J_j$. These are provided in appendix \ref{sec:potential_expansion}. 

Plugging the single-frequency profile $\f_\mathrm{osc}$ into the Kelin-Gordon equation and collecting the coefficient of $\cos(\omega t)$, we obtain the radial profile equation
\begin{align}\label{Minkowski_radial_eq}
(\nabla^2+\kappa_1^2) \phi_1(r) = J_1(r) ~,
\end{align}
where $\kappa_j^2\equiv (j\omega)^2-m^2$ is the square of the momentum for a particular mode and we have used $J_1(\f_1) = U_0'(\f_1)$. Oscillon profiles can be found by using the numerical shooting method and by demanding a localized, smooth and no-node solution for each $\omega$. Without loss of generality, we assume that the minimum of $V(\f)$ is located at $\f=0$ thus the boundary condition is $\phi_{1}(\infty)= 0$. Once the profile is found, the energy of oscillons can be obtained by time averaging the energy density over a period, and is given by
\begin{align} \label{Minkowski_energy}
E_{\mathrm{osc}} = \int_0^\infty \[ \frac{1}{4}\(\pd_r\phi_1 \)^2 + \frac{1}{4}(\omega^2 + m^2)\phi_1^2 + \frac{1}{2}U_0  \] 4\pi r^2dr~.
\end{align}
For these single frequency objects, we can define the particle number \cite{Mukaida:2016hwd}
\begin{align}
N_\mathrm{osc} = \frac{\omega}{2}\int_0^\infty \f_1^2 ~4\pi r^2 dr ~.
\end{align}
The stability condition of oscillons against small perturbations is given by \cite{Friedberg:1976me}
\begin{align}\label{stability_condition2}
\frac{dN_\mathrm{osc}}{d\omega} < 0 \quad\text{or}\quad \frac{dE_\mathrm{osc}}{d\omega} < 0~,
\end{align}
and the critical frequency $\omega_\mathrm{crit}$ can be obtained by setting $dE_\mathrm{osc}/d\omega = 0$.

Radial equations for radiating modes can be obtained by plugging the full expansion \eqref{profile_single_freq} into the Klein-Gordon equation and collecting the coefficient of $\cos(j\omega t)$, i.e.
\begin{align}\label{radial_radiation}
\[ \nabla^2 + \kappa_j^2 \]\xi_j(r) = S_j(r) \equiv J_j + \frac{1}{2}\sum_{k=3}^{\infty}\xi_k\( M_{|k+j|} + M_{|k-j|} \) ~,
\end{align}
where $j$ and $k$ are both odd, $S_j$ is the effective source and $M_j$ is the effective mass. For $\kappa_{j}^2>0$, the central amplitude of a radiating mode is
\begin{align}\label{xi_initial}
\xi_j(0) = - \int_{0}^{\infty} dr' ~S_j(r')~r'\cos(\kappa_j r') ~.
\end{align}
This is not in closed form, and we can find the solution of $\xi_j(r)$ by using the \emph{iterative method} developed in \cite{Zhang:2020bec}.\footnote{We will no longer call this method ``shooting'' as we did in \cite{Zhang:2020bec}, because technically we are not solving a boundary value problem.} The energy loss rate of oscillons is determined by radiation at large radius
\begin{align}
\frac{dE_\mathrm{osc}}{dt} = -4\pi r^2 \t{T}{^1_0} |_{r\rightarrow\infty} = \l. -4\pi r^2 \pd_t\xi\pd_r\xi \r|_{r\rightarrow\infty} ~,
\end{align}
where $\t{T}{^\mu_\nu} = \pd^\mu\f \pd_\nu\f + \delta^\mu_\nu \cal L$ is the energy-momentum tensor and the radiation is given by
\begin{align}\label{xi_inf}
\l. \xi(t,r) \r|_{r\rightarrow\infty} = -\frac{1}{4\pi r} \sum_{j=3}^\infty \til S_j(\kappa_j) \cos(\kappa_jr - j\omega t) ~.
\end{align}
Here $\til{S}_j(p)$ is the Fourier transform of the effective source
\begin{align}
\til S_j(p) = \int_0^\infty dr~ 4\pi r^2 \sinc(p r) S(r) ~.
\end{align}
The (absolute value of) decay rate is defined
\begin{align}\label{Gamma_Minkowski}
\Gamma \equiv \l| \la\dot E_\mathrm{osc}/E_\mathrm{osc} \ra_T \r|  = \frac{1}{8 \pi E_\mathrm{osc}} \sum_{j=3}^{\infty}\left[\tilde{S}_{j}(\kappa_{j})\right]^{2} j\omega~ \kappa_{j} \equiv \sum_{j=3}^{\infty} \Gamma_{j} ~,
\end{align}
where $\Gamma_j$ is the contribution due to $S_j$. Typically the radiating mode safisfies $|\xi_j|\gg|\xi_{j+2}|$, which means only finite terms are needed in the radial equation \eqref{radial_radiation}. However, the leading channel of decay rates $\Gamma_3$ might vanish for some $\omega$, causing a dip structure seen in figure \ref{fig:decayrateminkowski}. 

If we start with an oscillon with $\omega<\omega_{\rm dip}$, the frequency of the oscillons evolves to larger values by slowly emitting scalar radiation (and the profile changes correspondingly). Since the leading decay channel $\Gamma_3$ is vanishing in the dip, the oscillon configuration with $\omega\approx \omega_{\rm dip}$ is expected to have a long lifetime. When considering the total lifetime of oscillons that start out at $\omega< \omega_{\rm dip}$, the oscillon will spend most of its lifetime in such a dip. Generally speaking, we will use $\Gamma_3 + \Gamma_{5}$ to estimate the lifetime of oscillons, which is just the area enclosed by the evolution curve in $dt/dE_\mathrm{osc}$ versus $E_\mathrm{osc}$ plot, where $dt/dE_\mathrm{osc}=1/(\Gamma E_\mathrm{osc})$. For future reference, if we keep only $\xi_3$ and $\xi_5$, the effective source becomes
\begin{align}
\label{S3_Minkowski}
S_3 &= J_3 + \frac{1}{2}\xi_3 (M_0+M_6) + \frac{1}{2}\xi_5(M_2 + M_8) ~,\\
\label{S5_Minkowski}
S_5 &= J_5  + \frac{1}{2}\xi_3(M_2 + M_8) + \frac{1}{2}\xi_5 (M_0+M_{10}) ~.
\end{align}
The generalization to including more $\xi_j$ terms is straightforward. Note that the effective source $S_j$ receives a contribution from the oscillon background $J_j$ as well as corrections due to radiation $\xi_j$.

\subsection{A virial theorem and small parameters}
\label{sec:Minkowski_virial}
Assume that oscillons are single-frequency objects like Q-balls and Boson stars \cite{Lee:1991ax}, then one way to derive a virial theorem is to use the variational principle. The Legendre transformation
\begin{align}
F_\mathrm{osc} = \omega N_\mathrm{osc} - E_\mathrm{osc}
\end{align}
defines a functional of $\f_1$ and a function of $\omega$ (one may recognize that $F_\mathrm{osc}$ is just the Lagrangian). The variation of $F_\mathrm{osc}$ in terms of $\f_1$ by keeping $\omega$ fixed gives the profile equation of oscillons \eqref{Minkowski_radial_eq}. A virial theorem can be obtained by considering the variation $\f_1(r)\rightarrow \f_1(\lambda r)$ for an oscillons solution. By setting $\( \partial F_\mathrm{osc}/\pd \lambda \)_\omega = 0$
at $\lambda=1$, we find 
\begin{align}\label{virial_theorem}
E_\mathrm{S}/3 + E_\mathrm{V} = E_\mathrm{K} ~,
\end{align}
where the surface energy, potential energy and kinetic energy are defined
\begin{align}
E_\mathrm{S} = \int \frac{1}{4}(\pd_r\f_1)^2 d^3r \sep
E_\mathrm{V} = \int \( \frac{1}{4}m^2\f_1^2 + \frac{1}{2}U_0\) d^3r \sep
E_\mathrm{K} = \int \frac{1}{4}\omega^2\f_1^2 ~d^3r ~.
\end{align}

For oscillons with $\omega\lesssim m$, we can identify three small parameters immediately
\begin{align}\label{small_quantity}
\epsilon_r\equiv 1-\omega^2/m^2 \sep
\epsilon_V\sim \frac{U_0}{m^2\f_1^2/2} \sep
\epsilon_\xi \sim \frac{\xi_j}{\f_1} ~.
\end{align}
The parameter $\epsilon_r$ is a measure of how relativistic the particles inside the oscillon are. Equation \eqref{Minkowski_radial_eq} implies that at large radius $\f_1(r) \propto r^{-1} \exp\[ -(m^2-\omega^2)^{1/2}r \]$, hence a typical spatial derivative brings a factor $\sqrt{\epsilon_r} m$, that is, $\pd_r\f_1 \sim -\sqrt{\epsilon_r} m \f_1$, and
\begin{align}
\nabla^2\f_1\ll \omega^2\f_1 ~.
\end{align}
This means that the particles that make up the oscillon are \emph{non-relativistic}. And from equations \eqref{radial_radiation} and \eqref{virial_theorem}, we see $\epsilon_\xi\sim - \epsilon_V$ and $\epsilon_V\sim -\epsilon_r$.

For oscillons with $\omega \ll m$, we may not regard surface energy as a small quantity anymore. Take the tanh potential in \eqref{potentials} for example and assume a Gaussian profile
\begin{align}
\f_1(r) = C ~e^{-r^2/R^2} ~,
\end{align}
where $C\gg F$. Then each energy component becomes
\begin{align}
E_\mathrm{S} \sim C^2 R \sep
E_\mathrm{K} \sim C^2 R^3 \omega^2 \sim \frac{N_\mathrm{osc}^2}{C^2R^3}\sep
E_\mathrm{V} \sim R_\mathrm{V}^3 m^2 F^2 ~,
\end{align}
where we have taken advantage of the flatness of the potential at large $\f$ and $R_\mathrm{V}$ is the length scale satisfying $\f_1(R_\mathrm{V})\sim \cal O(1)$, i.e. 
\begin{align}
R_\mathrm{V}\sim R \log^{1/2} \( \frac{C}{F} \) \sim R ~.
\end{align}
By setting $\pd E_\mathrm{osc}/\pd R=\pd E_\mathrm{osc}/\pd C=0$ and keeping $N_\mathrm{osc}$ fixed, we obtain
\begin{align}
R\sim \omega^{-1}\sep
C\sim \omega^{-1} ~.
\end{align}
We see that three components of energy now are all comparable, and thus oscillon particles are relativistic. This phenomenon has been witnessed numerically in the context of dense axion stars \cite{Visinelli:2017ooc,Eby:2019ntd}.

\section{Full GR formalism}
\label{sec:GR}
So far our analysis does not include gravity. In this section, the scalar field is assumed to be minimally coupled to gravity. We introduce a Lagrangian mechanism that can convert the Hilbert action into one that contains only the first derivative of the metric in section \ref{sec:GR_Lagrangian}, then we use it to generalize the definition of energy and mass into curved spacetime in section \ref{sec:GR_definition}.

\subsection{A Lagrangian mechanism}
\label{sec:GR_Lagrangian}
The simplest choice of metric to describe oscillons is the spherical coordinates
\begin{align}\label{spherical_metric}
ds^2 = - e^{2\Phi(t,r)} \d t^2 + e^{-2\Psi(t,r)} \d r^2 + r^2 (\d\theta^2 + \sin^2\theta~ \d\j^2) ~,
\end{align}
where $\theta,\j$ are the polar and azimuthal angles, $r$ is $(2\pi)^{-1}$ times the circumference of a two-sphere. The action of our theory is composed of the action of gravity and matter $S= S_\mathrm{G} + S_\mathrm{M}$, specifically
\begin{align}
S_\mathrm{G} &= \frac{1}{16\pi G} \( \int_\Omega R \sqrt{-g}~ d^4x+ \int_{\pd\Omega}  K \sqrt{|h|}~ d^3x \) ~,\\
S_\mathrm{M} &= \int \[ -\frac{1}{2}g^{\mu\nu} \f_{,\mu}\f_{,\nu} - V(\f) \] \sqrt{-g} ~d^4x ~,
\end{align}
where $\f_{,\mu}\equiv \pd_\mu\f$ is defined for notation convenience, $R$ is the Ricci scalar and $g$ is the determinant of $g_{\mu\nu}$, i.e.
\begin{align}
R = & 2 r^{-2} - 2 e^{-2\Phi}\[ \Psi_{,00} - \Psi_{,0}(\Phi + \Psi )_{,0} \] \\
&- 2e^{2\Psi}\[ r^{-2} + 2 r^{-1}( \Phi + \Psi )_{,1} + \Phi_{,1}(\Phi + \Psi)_{,1} + \Phi_{,11} \] ~,\\
\sqrt{-g}~ d^4x =& e^{\Phi-\Psi} r^2 \sin\theta ~dt~dr~d\theta~d\j ~.
\end{align}
Apart from the standard Hilbert action, $K$ is a surface term and $h$ is the induced metric on the boundary $\pd\Omega$ \cite{Gibbons:1976ue} that can be appropriately chosen \cite{Lee:1988av}, i.e.
\begin{align}
K &= 2e^{\Psi} \Phi_{,1} + 4r^{-1} (e^\Psi - 1) ~,\\
\sqrt{|h|}~d^3x &= e^\Phi r^2 \sin\theta ~dt~d\theta~ d\j ~,
\end{align}
for the three dimensional surface at $r=r_0$ and
\begin{align}
K &= 2e^{-\Phi} \Psi_{,0} ~,\\
\sqrt{|h|}~d^3x &= e^{-\Psi} r^2 \sin\theta ~dr~d\theta~ d\j ~,
\end{align}
for that which is bounded by $t=\pm T$. The inclusion of surface terms will not change the Einstein equations, but can convert the Hilbert action into one that contains only the first derivative of the metric so that the usual Lagrangian mechanics can be applied. After setting $r_0$ and $T\rightarrow\infty$, the Lagrangian, i.e. $S = \int L ~dt$, becomes
\begin{align}
\label{Lagrangian_gravity}
L_\mathrm{G} &= (2G)^{-1} \int_0^\infty \[ e^{\Phi-\Psi} + e^{\Phi+\Psi}(1+2r\Phi_{,1}) - 2e^\Phi(1+r\Phi_{,1}) \] dr ~,\\
\label{Lagrangian_field}
L_\mathrm{M} &=  \int_0^\infty ( X-Y-V) ~e^{\Phi-\Psi} 4\pi r^2 dr ~,
\end{align}
where we have defined
\begin{align}
X \equiv \frac{1}{2}e^{-2\Phi}\f_{,0}^2 \sep 
Y \equiv \frac{1}{2}e^{2\Psi}\f_{,1}^2 ~.
\end{align}

The gravity and matter are related by the Einstein equation
\begin{align}
\t{G}{_\mu_\nu} = \t{R}{_\mu_\nu} - \frac{1}{2}g_{\mu\nu} R = 8\pi G ~\t{T}{_\mu_\nu} ~,
\end{align}
where $R_{\mu\nu}$ is the Ricci tensor and the energy-momentum tensor $\t T{^\mu_\nu}$ is given by
\begin{align}
T_{\mu\nu} = -2 \frac{1}{\sqrt{-g}} \frac{\delta S_\mathrm{M}}{\delta g^{\mu\nu}} = \f_{,\mu}\f_{,\nu} - \frac{1}{2}g_{\mu\nu} g^{\rho\sigma} \f_{,\rho}\f_{,\sigma} - g_{\mu\nu} V(\f) ~.
\end{align}
More specifically
\begin{align}
\label{G00_full}
\t{G}{^0_0} &= r^{-2}\( e^{2\Psi} - 1 + 2 r \Psi_{,1} e^{2\Psi}  \) = -8\pi G (X+Y+V) ~,\\
\label{G10_full}
\t{G}{^1_0} &= -2 r^{-1} \Psi_{,0} e^{2\Psi} = 8\pi G ~e^{2\Psi}\f_{,0}\f_{,1} ~,\\
\label{G11_full}
\t{G}{^1_1} &= r^{-2} \( e^{2\Psi} - 1 + 2 r \Phi_{,1} e^{2\Psi} \) = 8\pi G (X+Y-V) ~,\\
\label{G22_full}
\t{G}{^2_2}&= e^{-2\Phi} \[ \Psi_{,00} - \Psi_{,0} (\Phi + \Psi )_{,0} \] + e^{2\Psi} \[ \Phi_{,11} + (\Phi_{,1}+r^{-1}) (\Phi+\Psi )_{,1} \] = 8\pi G(X-Y-V) ~,
\end{align}
where $\t{G}{^3_3}=\t{G}{^2_2}$ and all other components vanish. The $\t{G}{^0_0}$ and $\t{G}{^1_1}$ equations can be alternatively obtained by varying the Lagrangian with respect to $\Phi$ and $\Psi$. The others can be derived using the contracted Bianchi identity $\t{G}{^\mu_{\nu;\mu}}=0$, i.e. $\nu=0$ gives \eqref{G10_full}, $\nu=2$ gives $\t{G}{^2_2}=\t{G}{^3_3}$ and $\nu=1$ gives \eqref{G22_full}. Some combanitions will be useful, for example, equations \eqref{G00_full} and \eqref{G11_full} give
\begin{align}\label{G11-G00}
r^{-1} e^{2\Psi} (\Phi-\Psi)_{,1} = 8\pi G(X+Y) ~,
\end{align}
and equations \eqref{G22_full} and \eqref{G11-G00} give
\begin{align}\label{G11+G22-G00}
e^{-2\Phi} \[ \Psi_{,00} - \Psi_{,0} (\Phi + \Psi )_{,0} \] + e^{2\Psi} \[ \nabla^2\Phi + \Phi_{,1} (\Phi+\Psi )_{,1} \] = 8\pi G(2X-V) ~,
\end{align}
where $\nabla^2\equiv \pd_r^2 + (2/r)\pd_r$. The equation of motion of $\f$ is
\begin{align}\label{EOM_phi_full_GR}
e^{-2\Phi}\[-\f_{,00} + (\Phi+\Psi)_{,0} \f_{,0} \] + e^{2\Psi} \[ \nabla^2\f +  (\Phi + \Psi)_{,1} \f_{,1} \] - V'(\f) = 0 ~,
\end{align}
which is obtained by varying the Lagrangian with respect to $\f$.

\subsection{Mass and energy}
\label{sec:GR_definition}
Following \cite{Lee:1988av}, we distinguish between the mass and energy of oscillons. At large radius the mass density vanishes exponentially, hence $\Phi$ and $\Psi$ scale as $r^{-1}$. The mass then must satisfy
\begin{align}\label{mass_Schwarzschild}
M_\mathrm{osc} = -G^{-1} \lim\limits_{r\rightarrow\infty} r\Psi = -G^{-1} \lim\limits_{r\rightarrow\infty} r\Phi ~,
\end{align}
to be consistent with the static Schwarzschild solution. There are other ways to express the same mass. For example, the LHS of 00 component of Einstein equation can be rewritten into $r^{-2} [ r\(e^{2\Psi}-1\) ]_{,1}$ hence the mass is also given by
\begin{align}\label{mass_T00}
M_\mathrm{osc} = \int_0^\infty ( X+Y+V) ~4\pi r^2 dr = -\int_{0}^{\infty} \t{T}{^0_0}~ 4\pi r^2 dr ~,
\end{align}
which is in agreement with the Schwarzschild mass \eqref{mass_Schwarzschild}.

A more enlightening way to describe the mass is to use the Hamiltonian formalism. The Lagrangian of matter \eqref{Lagrangian_field} indicates that the \emph{energy} of the oscillon is
\begin{align}\label{energy_oscillon_gravity}
E_\mathrm{osc} =  \int_0^\infty ( X+Y+V) ~e^{\Phi-\Psi} 4\pi r^2 dr = -\int \t{T}{^0_0} \sqrt{-g}~ d^3x  ~.
\end{align}
There is no kinetic term in the Lagrangian of gravity \eqref{Lagrangian_gravity} thus the energy of gravity is $E_\mathrm{G} = -L_\mathrm{G}$. Then we define the \emph{mass} of oscillons
\begin{align}
M_\mathrm{osc} \equiv E_\mathrm{osc} + E_G ~.
\end{align}
Combining the energy expression and the 00 component of Einstein equations, we find
\begin{align}
[r(e^\Phi-e^{\Phi+\Psi})]_{,1} = G (E_\mathrm{M} + E_\mathrm{G})_{,1} ~,
\end{align}
in agreement with the Schwarzschild mass \eqref{mass_Schwarzschild} and the ADM mass \cite{Arnowitt:1962hi}. After understanding what we mean by energy, the decay rate of oscillons can be studied numerically in appendix \ref{sec:GR_numerics} and analytically in the next section.

\section{Oscillons with linearized gravity}
\label{sec:linearized_gravity}
The typical central amplitude, radius and mass of dense oscillons in the non-relativistic limit are
\begin{align}
\f_1(0) \sim F \sep
R_\mathrm{osc} \sim 5 m^{-1} \sep
M_\mathrm{osc} \sim 100 F^2/m ~.
\end{align}
Nonlinear effects of gravitational interactions are not important if the size of oscillons is much smaller than their Schwarzschild radius
\begin{align}\label{epsilon_F}
R_\mathrm{osc} \ll GM_\mathrm{osc}  \Rightarrow
\epsilon_\phi \sim \f_1/M_\mathrm{pl} \ll 1  ~,
\end{align}
which is satisfied by a number of cosmological models. In this section, therefore, we study the decay rate and lifetime of oscillons in the non-relativistic limit and weak-field limit of gravity, specifically those with $\epsilon_r \lesssim 0.1$ and $\epsilon_\phi \lesssim 0.1$.\footnote{Examples of the effective field theory that focuses on such low-energy phenomena includes \cite{Mukaida:2016hwd, Eby:2018ufi, Braaten:2018lmj, Namjoo:2017nia, Salehian:2020bon}.} The basic idea is similar to what we have done in section \ref{sec:Minkowski}.

\subsection{Profiles}
In the weak-field approximation, the spherical metric \eqref{spherical_metric} reduces to
\begin{align}
ds^2 = -(1+2\Phi) \d t^2 + (1-2\Psi)\d r^2 + r^2(\d\theta^2 + \sin^2\theta ~\d\j^2) ~.
\end{align}
So far we have encountered five sets of small dimensionless parameters, recall equations \eqref{small_quantity} and \eqref{epsilon_F}, i.e. the spatial derivative parameter $\epsilon_r$, the nonlinear potential parameter $\epsilon_V$, the radiation parameter $\epsilon_\xi$, the amplitude parameter $\epsilon_\phi$ and the gravitational potentials $\Phi$ and $\Psi$ (denoted by $\epsilon_g$). To be consistent, we will keep all small quantities to 1st order, and to 2nd order if spatial derivatives of small parameters are involved. Then the equation of motion of $\f$ \eqref{EOM_phi_full_GR} becomes
\begin{align}\label{EOM_phi}
-(1-2\Phi)\f_{,00} + (\Phi+\Psi)_{,0}\f_{,0} + \nabla^2\f - V'(\f) = \cal O (\epsilon^2m^2\f) ~.
\end{align}

To be consistent with the field expansion \eqref{profile_single_freq}, we expand the gravitational potentials in terms of a Fourier cosine series
\begin{align}\label{expansion_Phi_Psi}
\Phi = \frac{1}{2}\Phi_0 + \sum_{j=2}^{\infty}\Phi_j\cos(j\omega t) \sep
\Psi = \frac{1}{2}\Psi_0 + \sum_{j=2}^{\infty}\Psi_j\cos(j\omega t) ~,
\end{align}
where $j$ is even. We can solve these radial modes by plugging the expansion into the Einstein equations and collecting the coefficient for each Fourier mode. Then equations \eqref{G10_full} and \eqref{G11_full} give
\begin{align}
\Psi_0 = - r\Phi_{0,1} + \cal O(\epsilon^2) \sep
\Psi_2 = \cal O(\epsilon^2) \sep
\Psi_{j\geq 4} = \cal O(\epsilon^3) ~,
\end{align}
and equations \eqref{G11-G00} and \eqref{G11+G22-G00} give
\begin{align}
\nabla^2\Phi_0 = 4\pi G ~m^2\f_1^2 + \cal O(\epsilon^3 m^2) \sep
\Phi_{2,1}/r = -2\pi G ~m^2\f_1^2 + \cal O(\epsilon^3 m^2) \sep
\Phi_{j\geq 4} =  \cal O(\epsilon^2)  ~.
\end{align}
Therefore, $-\Phi_0\sim -\Psi_0\sim \Phi_2\sim \epsilon_g\sim \epsilon_\f^2/\epsilon_r$.\footnote{For the quadratic potential, there is no mass scale $F$ and the central amplitude of oscillons satisfies $\phi_1\sim \cal O(\epsilon_r M_\mathrm{pl})$ \cite{Lee:1991ax}, hence $\epsilon_\f\sim \epsilon_g\sim \epsilon_r$.} Somewhat surprisingly, we find $\Phi_2$ and $\Psi_2$ are not the same order of magnitude, in constrast with the common results of the Newtonian gauge. We will ignore $\Phi_{j\geq 4}$ and $\Psi_{j\geq 2}$ in future calculations.

In order to find the profile of oscillons, we plug the field expansion \eqref{profile_single_freq} and \eqref{expansion_Phi_Psi} into equation \eqref{EOM_phi} and collect the coefficient of $\cos(\omega t)$ to get
\begin{align}\label{profile_gravity}
[ \nabla^2 + \omega^2(1-\Phi_0) -m^2 ]\f_1 = J_1 + \cal O(\epsilon^2m^2\f_1) ~.
\end{align}
Here $\omega(1-\Phi_0)^{1/2}$ can be regarded as an effective frequency and is larger than $\omega$. This equation can be solved by numerical shooting method.\footnote{The boundary conditions are $\f_1(\infty)\rightarrow 0$, $\Phi_0(\infty)\propto 1/r$ and $\Phi_2(\infty)\rightarrow 0$.} The time-averaged formula for the oscillon energy \eqref{energy_oscillon_gravity} is
\begin{align}
E_\mathrm{osc} = \int_0^\infty \[ \frac{1}{4}(\pd_r\f_1)^2 + \frac{1}{4}(\omega^2+m^2)\(1-\frac{1}{2}\Psi_0 + \frac{1}{2}\Phi_2 \) \f_1^2 + \frac{1}{2}U_0 \] 4\pi r^2dr + \cal O\(\frac{\epsilon^{1/2}\f_1^2}{m}\)~.
\end{align} 
As long as the oscillating part of the gravitational potentials is not too important, namely $\epsilon_g\lesssim 0.1$, oscillons share great similarities with mini-boson stars \cite{Friedberg:1986tp}, and we assume the stability condition is still valid \cite{Lee:1988av}
\begin{align}\label{stability}
\frac{dE_\mathrm{osc}}{d\omega} < 0 ~.
\end{align}
This is confirmed by comparing analytical predictions of $\omega_\mathrm{crit}$ with numerical results in the left panel of figure \ref{fig:lifetime}.

\subsection{Decay rates}
Plugging the field expansion \eqref{profile_single_freq} and \eqref{expansion_Phi_Psi} into equation \eqref{EOM_phi} and collecting the coefficient of $\cos(j\omega t)$, we obtain the radial equation of radiating modes
\begin{align}\label{radiation_eq}
\[\nabla^2 + \kappa_j^2\] \xi_j(r) = S_j(r) + \cal O(\epsilon^3m^2\f_1) ~,
\end{align}
where $j\geq 3$ and $j$ is odd, and $S_j$ is the effective source. We can keep finite terms of $\xi_j$ as we did in Minkowski spacetime. In particular, if we keep only $\xi_3$ and $\xi_5$, the effective source becomes
\begin{align}
\label{S3}
S_3 &= J_3 + \frac{1}{2}\xi_3(M_0+M_6) + \frac{1}{2}\xi_5(M_2+M_{8}) + 2 \omega^2\f_1 \Phi_2 + 9\omega^2\xi_3 \Phi_0 + 20\omega^2 \xi_5 \Phi_2~,\\
\label{S5}
S_5 &= J_5 + \frac{1}{2}\xi_3(M_2+M_8) + \frac{1}{2}\xi_5(M_0+M_{10}) + 12\omega^2\xi_3\Phi_2 + 25\omega^2 \xi_5 \Phi_0 ~,
\end{align}
where we have kept higher-order perturbations $\xi_5\Phi_j$ because of their large coefficients. Comparing \eqref{S3} with the corresponding expression in Minkowski spacetime \eqref{S3_Minkowski}, new corrections are introduced due to the coupling of gravity to oscillons and their radiation. The radiation equation can be solved by the iterative method \cite{Zhang:2020bec}.\footnote{The participation of $\Phi_0$ is possible to make the iteration divergent (since both $\Phi_0,\xi_3\propto 1/r$ at large radius), in which case $\xi_j$ can be easily found by adjusting initial values and matching the central amplitudes calculated by equations \eqref{xi_initial} and \eqref{radiation_eq}.}

The conservation law of the energy-momentum tensor is
\begin{align}
\t{T}{^{\mu}_{\nu;\mu}} = 0 \Rightarrow (\t{T}{^\mu_\nu} \sqrt{-g})_{,\mu} = 0 ~,
\end{align}
which implies
\begin{align}
\frac{dE_\mathrm{osc}}{dt} = \int  (\t{T}{^i_{0}}\sqrt{-g})_{,i} ~d^3x = \l. -4\pi r^2 \t{T}{^i_0} (1+\Phi-\Psi) \r|_{r\rightarrow\infty} = \l. -4 \pi r^2 \xi_{,0}\xi_{,1} \r|_{r\rightarrow\infty} ~, 
\end{align}
where we have used Gauss's divergence theorem to obtain the second equal sign. The $\xi$ at infinity is calculated in \eqref{xi_inf}, then the decay rate expression is just the same as the one in Minkowski spacetime \eqref{Gamma_Minkowski}.

\section{Examples}
\label{sec:examples}
In this section, we will explore gravitational effects on oscillon lifetimes by studying two examples, recall figure $\ref{fig:decayrateminkowski}$ and equation \eqref{potentials}, and comparing numerical results from full GR simulations with our analytical predictions. The numerical algorithm is described in appendix \ref{sec:GR_numerics}.

\subsection{The $\alpha$-attractor T-model of inflation}
The longevity of oscillons in this model (whose lifetime $\sim 10^{6}\,m^{-1}$ in Minkowski spacetime) is characterized by a dip structure in decay rates, see figure \ref{fig:decayrateminkowski}. In the dip, the leading channel of decay rates $\Gamma_3$ vanishes and the scalar radiation is dominated by the subleading channel $\Gamma_5$. Now we argue that the existence of gravity reduces the lifetime of oscillons.

\begin{figure}
	\centering
	\begin{minipage}{0.49\linewidth}
		\includegraphics[width=\linewidth]{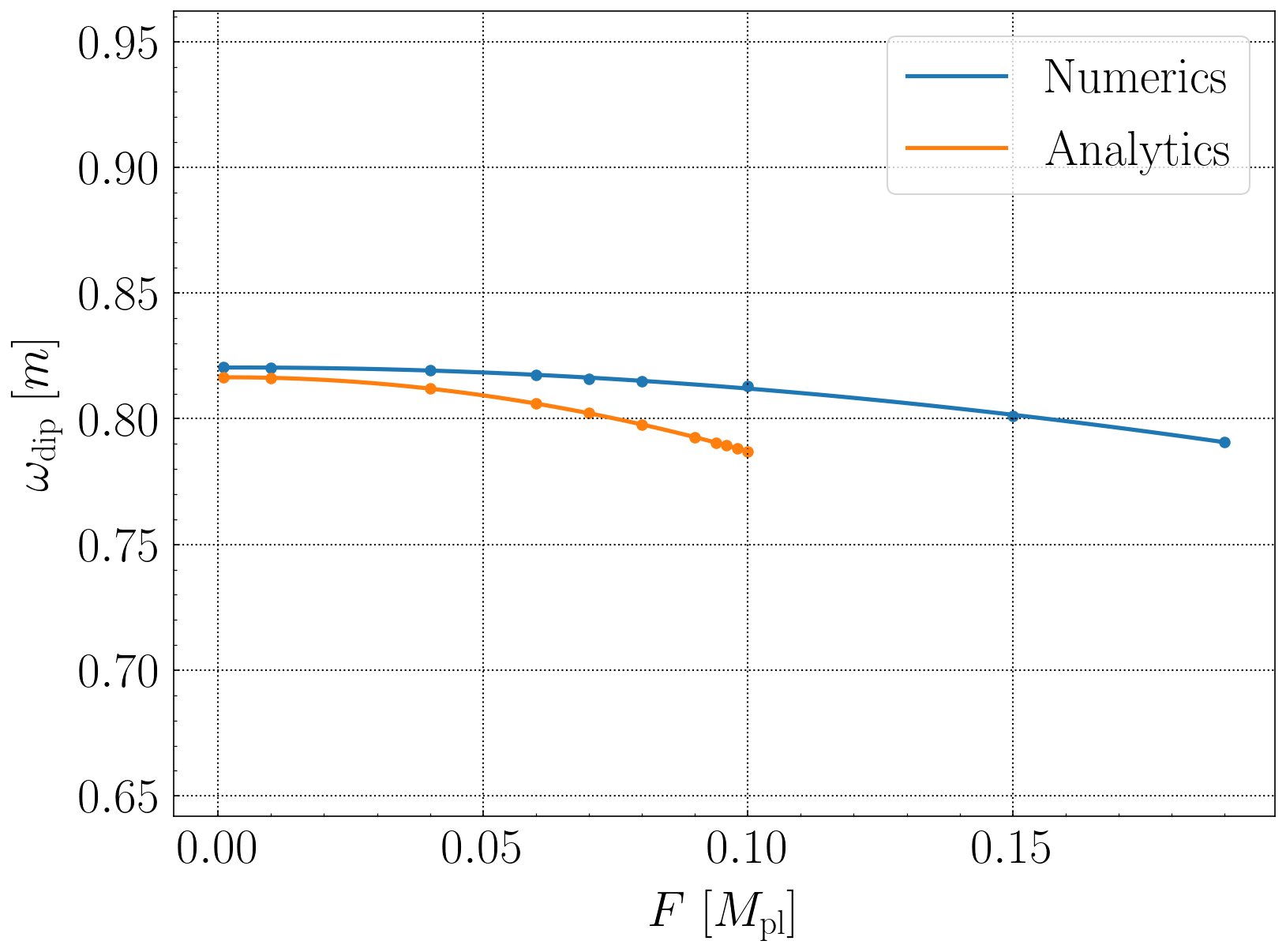}
	\end{minipage}
	\begin{minipage}{0.49\linewidth}
		\includegraphics[width=\linewidth]{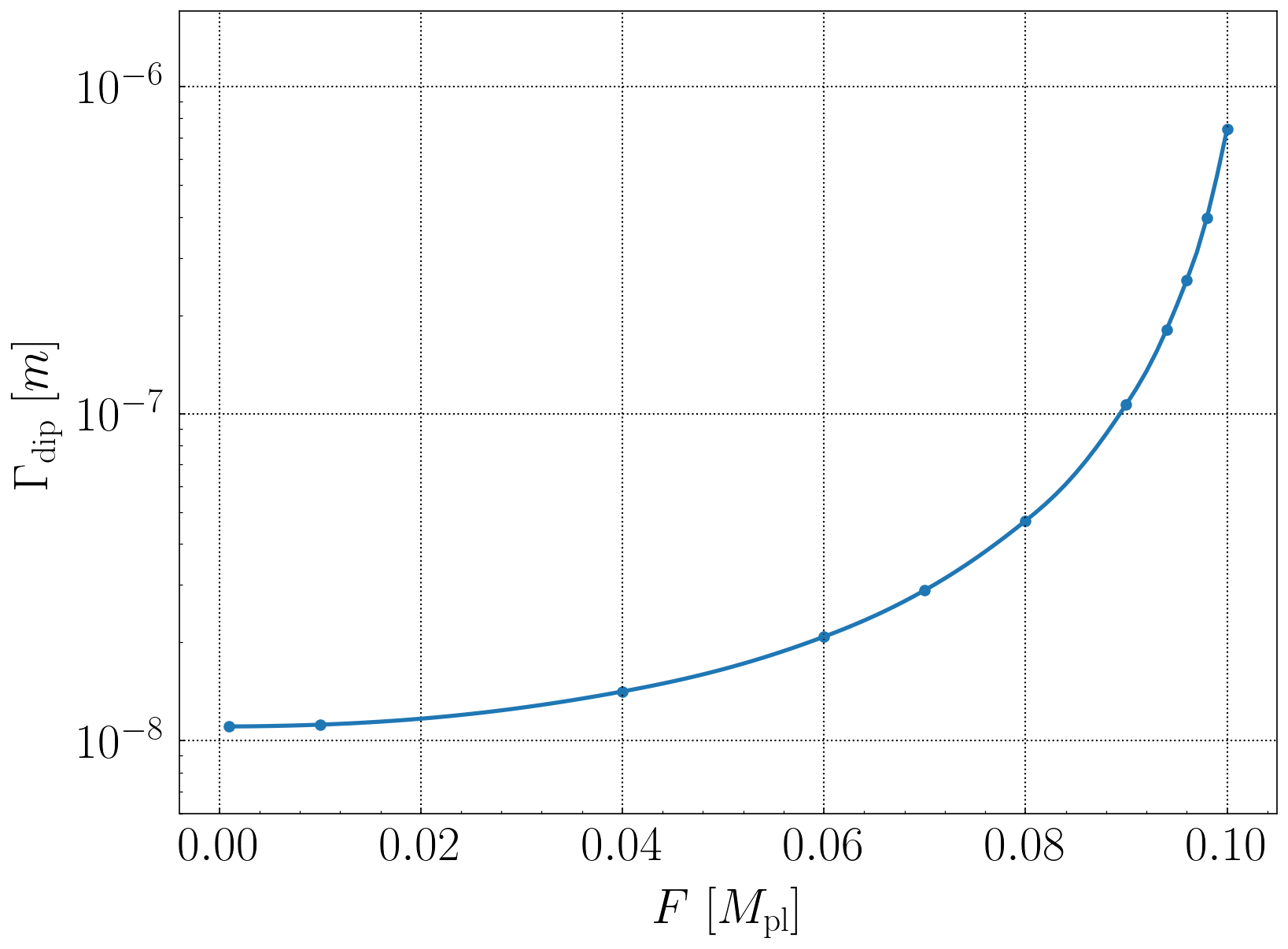}
	\end{minipage}
	\caption{\small Evidence that the existence of gravity reduces the lifetime of oscillons in the $\alpha$-attractor T-model. In the left panel, we show the dependence of dip locations on values of the characteristic mass scale $F$. The smaller value of $\omega_\mathrm{dip}$ is an indication of a shorter lifetime of oscillons, since the amplitude of radiating modes (in unit of $F$) is inversely related to frequencies and thus larger decay rates are expected. In the right panel, we confirm this expectation by explicitly calculating the decay rate at $\omega_\mathrm{dip}$.}
	\label{fig:tanh}
\end{figure}

We first study how the location of the dip is affected both analytically and numerically in figure \ref{fig:tanh} (left panel). The value of $\omega_\mathrm{dip}$ is significantly reduced when the mass scale $F$ approaches the reduced Planck mass. A smaller value of $\omega_\mathrm{dip}$ is an indication of shorter lifetimes, since the amplitude of radiating modes (in unit of $F$) is inversely related to frequencies and thus larger decay rates are expected. The comparison between analytics and numerics also provides a chance to test our formalism, which correctly captures the gravitational effect on $\omega_\mathrm{dip}$ as long as the assumptions of weak-field gravity and non-relativistic limit remain valid.

\begin{figure}
	\centering
	\includegraphics[width=0.5\linewidth]{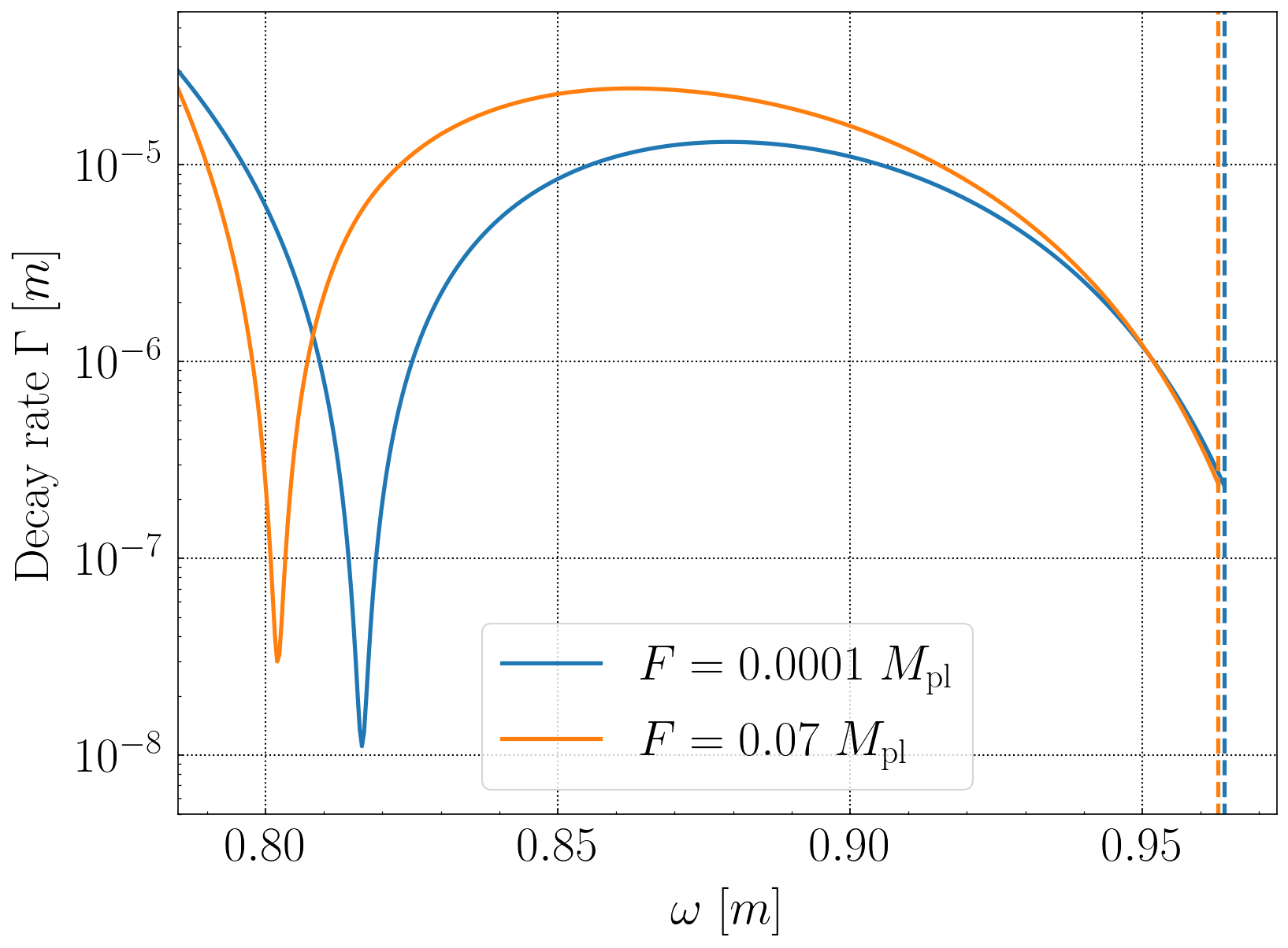}
	\caption{\small A comparison of decay rates for two different values of $F$ in the $\alpha$-attractor T-model.}
	\label{fig:tanhdecayrate}
\end{figure}

To exclude the possibility that the subleading channel of decay rates also vanishes, we explicitly calculate $\Gamma_5$ around $\omega_\mathrm{dip}$ as shown in figure \ref{fig:tanh} (right panel). As a determinant factor, the increasing of $\Gamma_\mathrm{dip}\approx \Gamma_5$ is a clear evidence that the existence of gravity reduces the lifetime of oscillons, which has also been confirmed numerically. For convenience, we present a direct visualization of these two factors in figure \ref{fig:tanhdecayrate}.\footnote{We thank Andrew Long for a suggestion of making this plot.}

\subsection{The axion monodromy model}
In constrast with the $\alpha$-attractor T-model, the subleading channel of decay rates $\Gamma_5$ is never comparable with the leading channel $\Gamma_3$ in the axion monodromy model. The longevity of oscillons in this case (whose lifetime $\sim 10^{8}\,m^{-1}$ in Minkowski spacetime) is due to the dramatic suppresion of $\Gamma_3$ just before their final collapse at $\omega_\mathrm{crit}$. Consequently, there are two factors that determine the lifetime of oscillons, the value of $\omega_\mathrm{crit}$ and the decay rate around $\omega_\mathrm{crit}$.

\begin{figure}
	\centering
	\begin{minipage}{0.485\linewidth}
		\includegraphics[width=\linewidth]{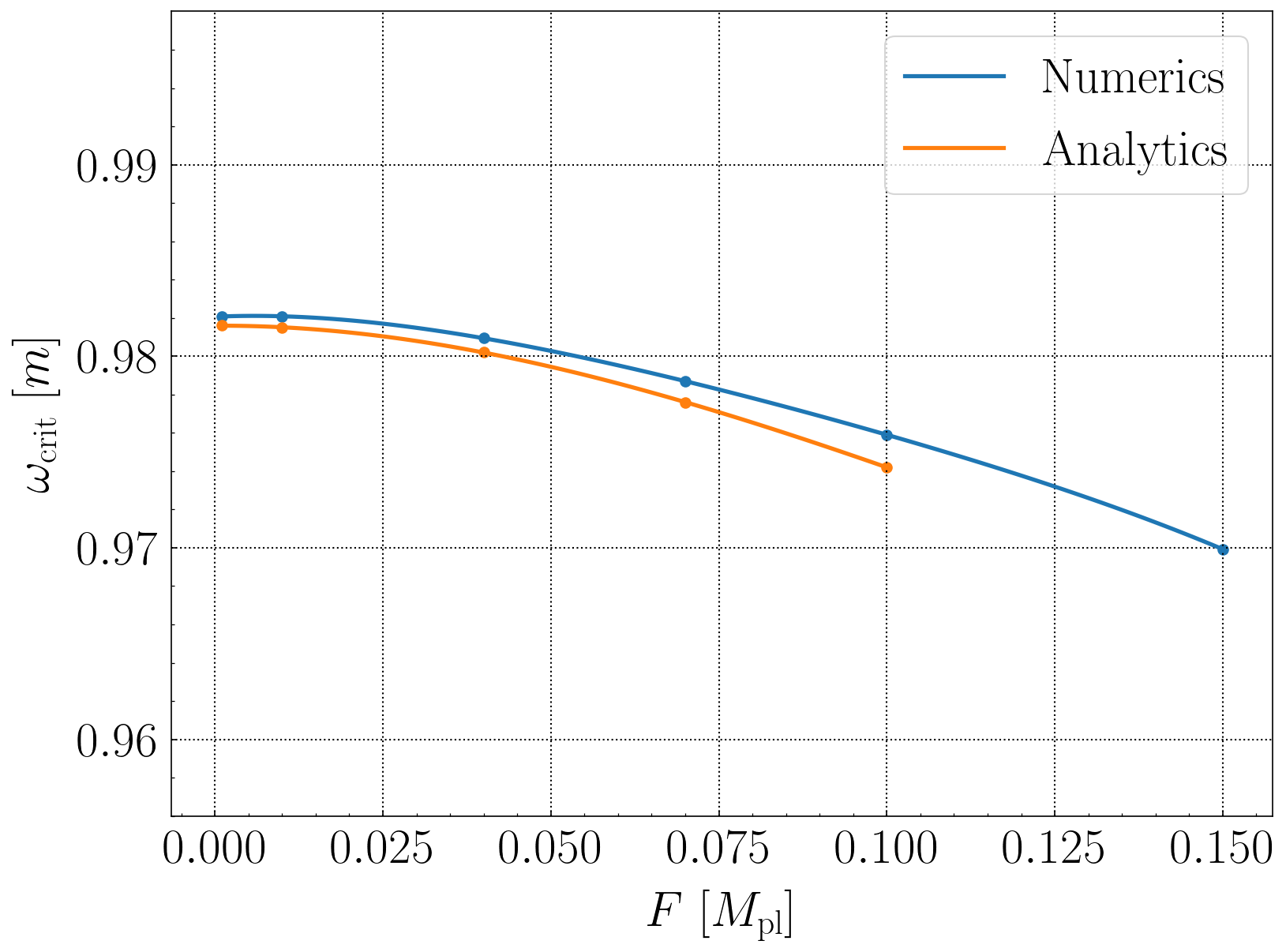}
	\end{minipage}
	\begin{minipage}{0.495\linewidth}
		\includegraphics[width=\linewidth]{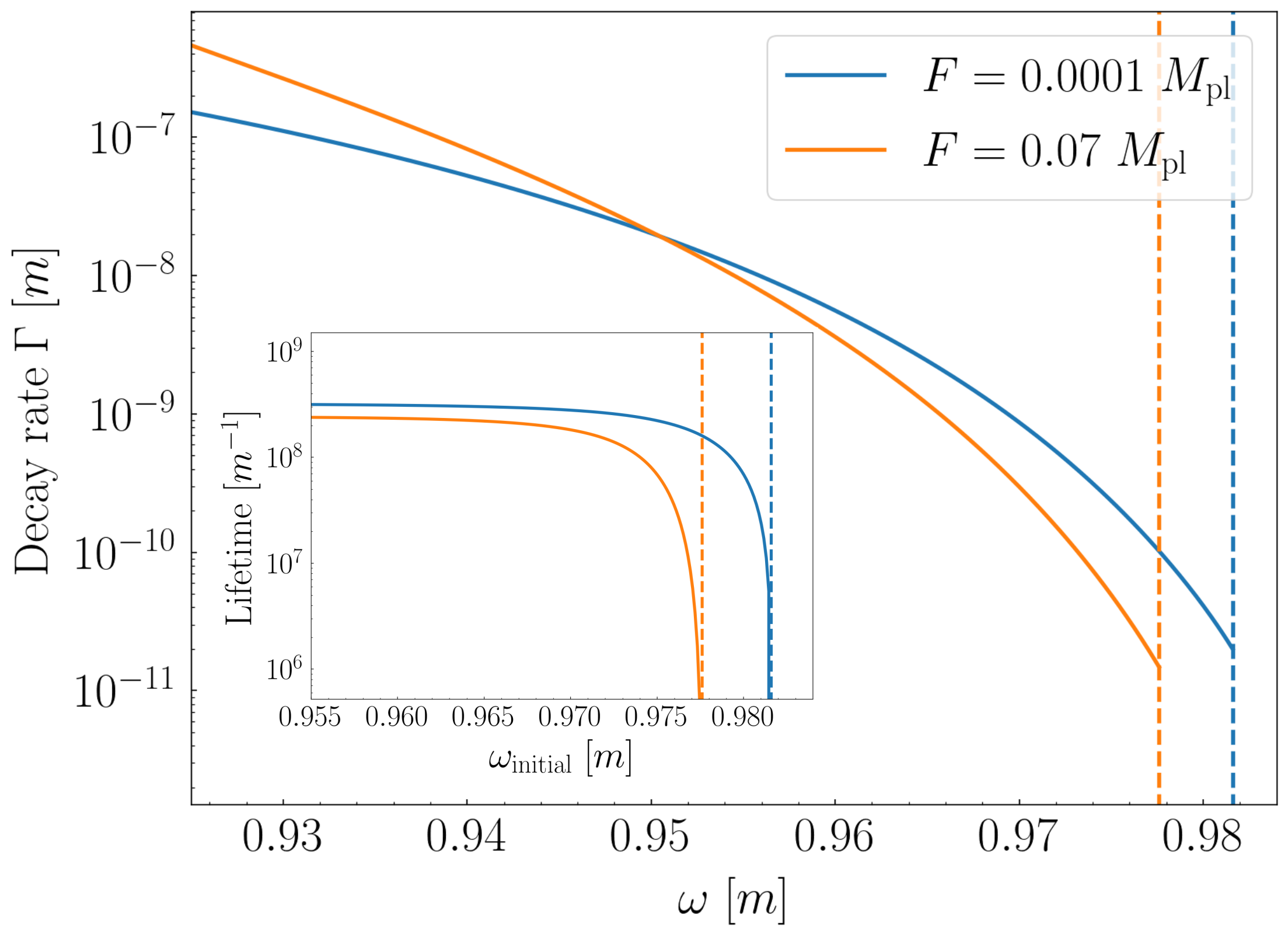}
	\end{minipage}
	\caption{\small Two determinant factors of oscillon lifetimes in the axion monodromy model, values of the critical frequency (left panel) and decay rates around $\omega_\mathrm{crit}$ (right panel). It is shown that gravitational effects decrease the value of $\omega_\mathrm{crit}$, which tends to reduce the lifetime, while suppressing the decay rate around $\omega_\mathrm{crit}$, which tends to stablize oscillons. In the inset, we show that oscillon lifetimes are reduced slightly for the stronger-field gravity.}
	\label{fig:lifetime}
\end{figure}

In figure \ref{fig:lifetime} (left panel), we show that the values of $\omega_\mathrm{crit}$ are inversely related to $F$. This seems an indication of shorter lifetimes for stronger gravitational effects because oscillons now spend less time around $\omega_\mathrm{crit}$. A good match between analytics and numerics implies that the stability condition \eqref{stability} is still valid as long as gravity is not too important.

Based on our semi-analytical framework, we calculate oscillon decay rates for $F=0.0001 M_\mathrm{pl}$ and $F=0.07 M_\mathrm{pl}$ in the right panel. Compared with the very weak-field gravity, it is shown that the decay rate around $\omega_\mathrm{crit}$ is more suppressed for $F=0.07 M_\mathrm{pl}$. This can be qualitatively understood by inspecting equations \eqref{S3_Minkowski} and \eqref{S3}. Since $J_3$ at $\omega_\mathrm{crit}$ typically has a Gaussian-like shape with a negative amplitude, the introduction of the positive term $2\omega_2\f_1\Phi_2$ diminishes the magnitude of the effective source, and thus tends to reduce the decay rate and increase the total lifetime.

To determine which factor dominates, we integrate out the decay rate and show the lifetime in an inset of the right panel of figure \ref{fig:lifetime}. Our results imply that the first factor plays a more important role and the oscillon lifetime is shorter for stronger-field gravity. Nevertheless, the oscillon is still too long-lived to be simulated with our current numerical algorithms. And we leave this as a testable prediction for a future numerical experiment.

\section{Conclusions}
\label{sec:summary}
In this paper we generalize the method developed in our previous work \cite{Zhang:2020bec} to include gravitational interactions, by expanding equations in terms of five small parameters
\begin{align}
\epsilon_r \sim \frac{\pd_r^2}{m^2} \sep
\epsilon_V \sim \frac{V_\mathrm{nl}}{m^2\f_\mathrm{osc}^2} \sep
\epsilon_\xi \sim \frac{\xi}{\f_\mathrm{osc}} \sep
\epsilon_\f \sim \frac{\f_\mathrm{osc}}{M_\mathrm{pl}} \sep
\epsilon_g \sim \Phi ~,
\end{align}
where $\epsilon_r\sim -\epsilon_V \sim \epsilon_\xi$ and $\epsilon_g\sim \epsilon_\f^2/\epsilon_r$ for dense oscillons. We have shown in \cite{Zhang:2020bec} that keeping just the leading order of $\epsilon_r, \epsilon_V$ and $\epsilon_\xi$ gives accurate predictions of oscillon decay rates and lifetimes in Minkowski spacetime. Thus in this paper, we explore the impacts of $\epsilon_\f$ and $\epsilon_g$ by keeping all other parameters fixed. The application of our semi-analytical and perturbative framework is based on two assumptions, the non-relativistic limit of oscillons and the weak-field limit of gravity, namely $\epsilon_r\lesssim 0.1$ and $\epsilon_\f\lesssim 0.1$. 

In order to explore the gravitational effects on oscillon lifetimes, we discuss two well-motivated and representative models in detail, the $\alpha$-attractor T-model of inflation and the axion monodromy model with potentials given by \eqref{potentials}. As presented in figure \ref{fig:tanh}, \ref{fig:tanhdecayrate} and \ref{fig:lifetime}, our results for $F\sim 0.1M_\mathrm{pl}$ show the following:
\begin{itemize}
	\item In the $\alpha$-attractor T-model, oscillon lifetimes are dominated by an exceptionally stable intermediate configuration, which is visualized as a dip structure in figure \ref{fig:decayrateminkowski}. We show that the existence of gravity decreases the value of $\omega_{\rm dip}$ and increases the decay rate at $\omega_{\rm dip}$. As a result, oscillon lifetimes are reduced when gravity becomes more important.
	\item In the axion monodromy model, oscillons are the most stable just before their final collapse at $\omega_\mathrm{crit}$. It is shown that stronger gravitational effects suppress the decay rate around $\omega_\mathrm{crit}$, which tends to stablize oscillons, while diminishing the value of $\omega_\mathrm{crit}$, which tends to decrease the lifetime. By explicitly integrating out the decay rate, we find that the latter factor dominates and oscillon lifetimes are reduced slightly.
\end{itemize}
In both examples we have considered, the evolution of an oscillon is almost identical to that in Minkowski spacetime if $F\ll 0.1 M_\mathrm{pl}$. Therefore, in such cases, one may study the decay rate and dynamics of at least single oscillons by ignoring gravity.

For stronger-field gravity, i.e. $\epsilon_\f\gtrsim 0.1$ or $F\gtrsim 0.1 M_\mathrm{pl}$, our equations with leading-order corrections are no longer accurate. Also note that both $\epsilon_\f$ and $\epsilon_g$ increase rapidly when we probe smaller frequences. Thus one should keep higher-order perturbations of $\epsilon_\f$ and $\epsilon_g$ to obtain more reliable conclusions. But in this regime, exotic phenomena such as black hole formation \cite{Helfer:2016ljl, Muia:2019coe, Nazari:2020fmk} might be more interesting than oscillon lifetimes. In future work, we will further show that a new phenomenon of migration from dense to dilute oscillons makes the notion of oscillon lifetimes less meaningful.

\acknowledgments

We would especially like to thank Mustafa Amin for many stimulating discussions, initial collaboration and insightful advice. We would like to thank Paul Saffin for a careful reading of the manuscript and suggestions for its improvement, and Borna Salehian for helpful discussions on scalar perturbation theory. We also thank Mudit Jain, Andrew Long, Kaloian Lozanov and Zong-Gang Mou for useful comments. This work is supported by a NASA ATP theory grant NASA-ATP Grant No. 80NSSC20K0518.

\appendix
\section{Fourier cosine coefficients of scalar potentials} \label{sec:potential_expansion}
In order to build some intuition of how $U_j,J_j$ and $M_j$ behave, let us consider a symmetric polynomial potential of the general form
\begin{align}\label{polynomial_potential}
V_\mathrm{nl}(\f) = \sum_{n=4}^{\infty} \frac{g_n}{n!} \f^n ~,
\end{align}
where $n$ is even. By using the oscillon profile $\f_\mathrm{osc} = \f_1 \cos(\omega t)$ and the identity
\begin{align}
\cos^n(\omega t) = \sum_{k=0}^{n} \frac{1}{2^n} \frac{n!}{(n-k)!k!} \cos[ (n-2k) \omega t ] ~,
\end{align}
we find the potentials can be written in terms of Fourier cosine series
\begin{align}\label{potential_U}
U = &
\sum_{n=4}^{\infty} \frac{g_n}{2^n \( \frac{n}{2} ! \)^2} \f_1^n 
+ \sum_{n=4}^{\infty} \frac{g_n}{2^{n-1} \( \frac{n+2}{2} \)! \( \frac{n-2}{2} \)!} \f_1^n \cos(2\omega t) \\
&+ \sum_{j=4}^{\infty} \sum_{n=j}^{\infty} \frac{g_n}{2^{n-1} \( \frac{n+j}{2} \)! \( \frac{n-j}{2} \)!} \f_1^n \cos(j\omega t) ~,\\
M = &
\sum_{n=2}^\infty \frac{g_{n+2}}{2^n\( \frac{n}{2}! \)^2} \f_1^n
+ \sum_{j=2}^{\infty} \sum_{n=j}^{\infty} \frac{g_{n+2}}{2^{n-1} \( \frac{n+j}{2} \)! \( \frac{n-j}{2} \)!} \f_1^n \cos(j\omega t) ~,
\end{align}
where $j$ and $n$ are even, and
\begin{align}
J = \sum_{n=3}^\infty \frac{g_{n+1}}{2^{n-1} \( \frac{n+1}{2} \)! \( \frac{n-1}{2} \)!} \f_1^n \cos(\omega t) 
+ \sum_{j=3}^\infty \sum_{n=j}^{\infty} \frac{g_{n+1}}{2^{n-1} \( \frac{n+j}{2} \)! \( \frac{n-j}{2} \)!} \f_1^n \cos(j\omega t) ~,
\end{align}
where $j$ and $n$ are odd.

\section{Numerical algorithms}
\label{sec:GR_numerics}
In this appendix we introduce briefly our numerical algorithm. For convenience we will set $M_\mathrm{pl}\equiv 1$ so that all the mass is in unit of reduced Planck mass. And to make the equations more accessible to coding, define
\begin{align}
A\equiv e^{-\Psi} \sep
B\equiv e^{\Phi+\Psi} \sep
v \equiv B^{-1}\f_{,0} \sep
u \equiv \f_{,1} ~.
\end{align}
Then the equation of motion of $\f$ becomes
\begin{align}\label{numerical_equations}
\f_{,0} = v B \sep
u_{,0} = (vB)_{,1} \sep
v_{,0} = r^{-2} (r^2 B u)_{,1} - A^2 B V'(\f) ~,
\end{align}
where the metric can be given by 00 and 11 components of Einstein equations, i.e.
\begin{align}
\label{numerics_A}
A_{,1} &= \frac{A}{2}\[ \frac{1-A^2}{r} + \frac{r}{2} \( v^2 + u^2 + 2A^2V \) \] ~,\\
\label{numerics_B}
B_{,1} &= B\( \frac{A^2-1}{r} - rA^2V \) ~.
\end{align}
One is also recommended to rescale the fields in unit of $F$ to reduce the roundoff errors brought by small numbers when $F\ll M_\mathrm{pl}$. The oscillon energy can be calculated by
\begin{align}
E_\mathrm{osc} = \int_0^\infty \frac{1}{2}B(v^2+u^2+2A^2 V) ~4\pi r^2 dr ~.
\end{align}

To maintain the smoothness at the center $r=0$, we must require
\begin{align}
u=v_{,1}=0 \sep
A = 1 \sep
A_{,1} = B_{,1} = 0 ~.
\end{align}
To appropriately account for the origin, we set a spatial grid $r_n = (n-1/2)dr$ with a fictitious point $r_0=-dr/2$. The inner boundary conditions then become a parity condition: $\f,v,A,B$ are even and $u$ is odd. For the outer boundary conditions we adopt the radiative boundary conditions \cite{Alcubierre:2000xu} by assuming that the dynamical variables $\f$ and $v$ behave like spherical waves $\f(t,r) = f(r-t)/r$. In practice, we will use this in the differential form
\begin{align}\label{numerics_boundary}
\phi_{,0} + \phi_{,1} + \phi/r = 0 \sep
v_{,0} + v_{,1} + v/r = 0 \sep
v + u + \f/r = 0 ~,
\end{align}
for $\phi,v,u$ repectively. Note that one actually does not need boundary conditions for $\phi$, but we find that integrating $\phi$ all the way to the boundary point will inevitably generate instabilities in long-time simulations. Finally, we note that the Schwarzchild metric should be recovered at large $r$ hence $A$ and $B$ satisfy 
\begin{align}\label{initial_B}
B(r\rightarrow\infty) = 1/A^2(r \rightarrow\infty) ~.
\end{align}

We adopt 4th-order Runge-Kutta method as the time integrator to evolve equations \eqref{numerical_equations}, while the spatial derivatives are discretized by the standard 4th-order centered difference \cite{Zlochower:2005bj}. The 6th-order Kreiss-Oliger dissipation with strength parameter $\epsilon=0.001$ \cite{alcubierre2008introduction} is added except boundary points to avoid shock waves while the accuracy is still remained. The boundary condition \eqref{numerics_boundary} is discretized by finite difference methods, specifically a 2nd-order upwind method \cite{alcubierre2008introduction}
\begin{align}
v^{m+1}_n = \frac{1}{1+dt/dr} \[ \( 1-\frac{dt}{dr}-\frac{2dt}{r_n} \) v^m_n + \(1+\frac{dt}{dr} \) v^m_{n-1} - \(1-\frac{dt}{dr}\) v^{m+1}_{n-1} \] ~,
\end{align}
where $m,n$ are the temporal and spatial indices respectively. Once $\phi,u,v$ have been advanced for one time level, we use the 4th-order Runge-Kutta method to integrate \eqref{numerics_A} outwards and \eqref{numerics_B} inwards to get $A,B$. The initial values of $B$ are obtained by using the Schwarzchild condition \eqref{initial_B}.

We typically set the boundary at a finite value $r_f$, which should be significantly larger than the oscillon size, and $dr=2dt$ to satisfy the Courant-Friedrichs-Lewy condition. All the physical quantities are averaged over a time window $T_\mathrm{ave}=300 ~\mathrm{m^{-1}}$, which is much larger than one period $2\pi/\omega$ but smaller than oscillon lifetime, unless otherwise stated. We have checked that slight changes of parameters (i.e. $dt,dr,r_f,\epsilon,T_\mathrm{ave}$) do not affect the results significantly.

\begin{figure}
	\centering
	\includegraphics[width=0.5\linewidth]{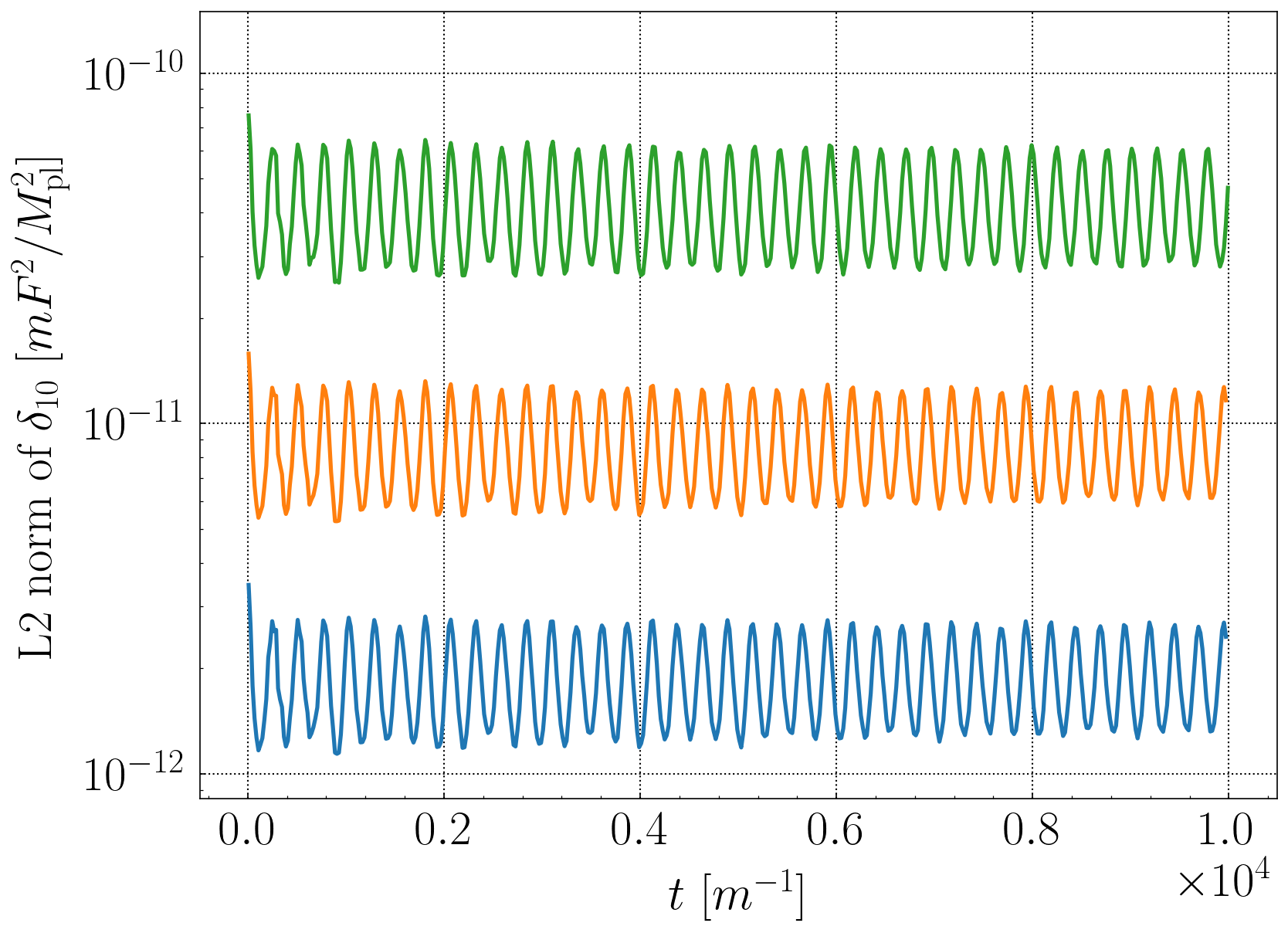}
	\caption{\small Convergence test for axion monodromy-type potential with $F=10^{-3} M_\mathrm{pl}$. Three colors refer to three different resolutions: $dt=0.025 m^{-1}$ (blue), $dt=0.05 m^{-1}$ (orange) and $dt=0.1 m^{-1}$ (green). The initial conditions correspond to a Gaussian profile $\f(r)=1.43 F \exp(-m^2 r^2/5.2^2)$. And we have set $T_\mathrm{ave}=20 m^{-1}$ to average the data. The fact that the L2 norm values increase by a factor of 4 every time we double the time step shows that the code maintains 2nd-order convergent.}
	\label{fig:codetest}
\end{figure}

Although we tried to implement 4th-order methods to evolve the system, the $1/r$ terms at small $r$ and the boundary discretization will destroy the 4th-order accuracy, hence we expect our algorithm to maintain 2nd-order. The codes can be tested by an exact equation
\begin{align}
\delta_{10} \equiv A_{,0} - rABuv/2 = 0 ~,
\end{align}
which is just the $10$ component of Einstein equations. As an example, we plot L2 norm of $\delta_{10}$ across the grids in figure \ref{fig:codetest}, where the L2 norm is defined by $||\f||_2 \equiv \sqrt{ \frac{1}{N}  \sum_{n=0}^{N-1}\f_n^2  }$ ,
and $\f_n$ denotes the value of $\f$ at the spatial point $r_n$.

\bibliographystyle{JHEP}
\bibliography{reference}

\providecommand{\href}[2]{#2}\begingroup\raggedright\begin{thebibliography}{10}

\bibitem{osti_4051808}
A.~E. Kudryavtsev, \emph{Solitonlike solutions for a higgs scalar field},
  {\emph{JETP Lett. (USSR) (Engl. Transl.), v. 22, no. 3, pp. 82-83} }.

\bibitem{Bogolyubsky:1976yu}
I.~Bogolyubsky and V.~Makhankov, \emph{{Lifetime of Pulsating Solitons in Some
  Classical Models}}, {\emph{Pisma Zh. Eksp. Teor. Fiz.} {\bfseries 24} (1976)
  15}.

\bibitem{Copeland:1995fq}
E.~J. Copeland, M.~Gleiser and H.-R. Muller, \emph{{Oscillons: Resonant
  configurations during bubble collapse}},
  \href{https://doi.org/10.1103/PhysRevD.52.1920}{\emph{Phys. Rev. D}
  {\bfseries 52} (1995) 1920}
  [\href{https://arxiv.org/abs/hep-ph/9503217}{{\ttfamily hep-ph/9503217}}].

\bibitem{Amin:2010jq}
M.~A. Amin and D.~Shirokoff, \emph{{Flat-top oscillons in an expanding
  universe}}, \href{https://doi.org/10.1103/PhysRevD.81.085045}{\emph{Phys.
  Rev.} {\bfseries D81} (2010) 085045}
  [\href{https://arxiv.org/abs/1002.3380}{{\ttfamily 1002.3380}}].

\bibitem{Amin:2013ika}
M.~A. Amin, \emph{{K-oscillons: Oscillons with noncanonical kinetic terms}},
  \href{https://doi.org/10.1103/PhysRevD.87.123505}{\emph{Phys. Rev.}
  {\bfseries D87} (2013) 123505}
  [\href{https://arxiv.org/abs/1303.1102}{{\ttfamily 1303.1102}}].

\bibitem{Seidel:1993zk}
E.~Seidel and W.-M. Suen, \emph{{Formation of solitonic stars through
  gravitational cooling}},
  \href{https://doi.org/10.1103/PhysRevLett.72.2516}{\emph{Phys. Rev. Lett.}
  {\bfseries 72} (1994) 2516}
  [\href{https://arxiv.org/abs/gr-qc/9309015}{{\ttfamily gr-qc/9309015}}].

\bibitem{Farhi:2007wj}
E.~Farhi, N.~Graham, A.~H. Guth, N.~Iqbal, R.~Rosales and N.~Stamatopoulos,
  \emph{{Emergence of Oscillons in an Expanding Background}},
  \href{https://doi.org/10.1103/PhysRevD.77.085019}{\emph{Phys. Rev. D}
  {\bfseries 77} (2008) 085019}
  [\href{https://arxiv.org/abs/0712.3034}{{\ttfamily 0712.3034}}].

\bibitem{Amin:2010xe}
M.~A. Amin, \emph{{Inflaton fragmentation: Emergence of pseudo-stable inflaton
  lumps (oscillons) after inflation}},
  \href{https://arxiv.org/abs/1006.3075}{{\ttfamily 1006.3075}}.

\bibitem{Levkov:2018kau}
D.~Levkov, A.~Panin and I.~Tkachev, \emph{{Gravitational Bose-Einstein
  condensation in the kinetic regime}},
  \href{https://doi.org/10.1103/PhysRevLett.121.151301}{\emph{Phys. Rev. Lett.}
  {\bfseries 121} (2018) 151301}
  [\href{https://arxiv.org/abs/1804.05857}{{\ttfamily 1804.05857}}].

\bibitem{Amin:2010dc}
M.~A. Amin, R.~Easther and H.~Finkel, \emph{{Inflaton Fragmentation and
  Oscillon Formation in Three Dimensions}},
  \href{https://doi.org/10.1088/1475-7516/2010/12/001}{\emph{JCAP} {\bfseries
  1012} (2010) 001} [\href{https://arxiv.org/abs/1009.2505}{{\ttfamily
  1009.2505}}].

\bibitem{Amin:2011hj}
M.~A. Amin, R.~Easther, H.~Finkel, R.~Flauger and M.~P. Hertzberg,
  \emph{{Oscillons After Inflation}},
  \href{https://doi.org/10.1103/PhysRevLett.108.241302}{\emph{Phys. Rev. Lett.}
  {\bfseries 108} (2012) 241302}
  [\href{https://arxiv.org/abs/1106.3335}{{\ttfamily 1106.3335}}].

\bibitem{Lozanov:2014zfa}
K.~D. Lozanov and M.~A. Amin, \emph{{End of inflation, oscillons, and
  matter-antimatter asymmetry}},
  \href{https://doi.org/10.1103/PhysRevD.90.083528}{\emph{Phys. Rev. D}
  {\bfseries 90} (2014) 083528}
  [\href{https://arxiv.org/abs/1408.1811}{{\ttfamily 1408.1811}}].

\bibitem{Lozanov:2016hid}
K.~D. Lozanov and M.~A. Amin, \emph{{Equation of State and Duration to
  Radiation Domination after Inflation}},
  \href{https://doi.org/10.1103/PhysRevLett.119.061301}{\emph{Phys. Rev. Lett.}
  {\bfseries 119} (2017) 061301}
  [\href{https://arxiv.org/abs/1608.01213}{{\ttfamily 1608.01213}}].

\bibitem{Lozanov:2017hjm}
K.~D. Lozanov and M.~A. Amin, \emph{{Self-resonance after inflation: oscillons,
  transients and radiation domination}},
  \href{https://doi.org/10.1103/PhysRevD.97.023533}{\emph{Phys. Rev.}
  {\bfseries D97} (2018) 023533}
  [\href{https://arxiv.org/abs/1710.06851}{{\ttfamily 1710.06851}}].

\bibitem{Kou:2019bbc}
X.-X. Kou, C.~Tian and S.-Y. Zhou, \emph{{Oscillon Preheating in Full General
  Relativity}},  \href{https://arxiv.org/abs/1912.09658}{{\ttfamily
  1912.09658}}.

\bibitem{Olle:2019kbo}
J.~Ollé, O.~Pujolàs and F.~Rompineve, \emph{{Oscillons and Dark Matter}},
  \href{https://arxiv.org/abs/1906.06352}{{\ttfamily 1906.06352}}.

\bibitem{Amin:2019ums}
M.~A. Amin and P.~Mocz, \emph{{Formation, gravitational clustering, and
  interactions of nonrelativistic solitons in an expanding universe}},
  \href{https://doi.org/10.1103/PhysRevD.100.063507}{\emph{Phys. Rev. D}
  {\bfseries 100} (2019) 063507}
  [\href{https://arxiv.org/abs/1902.07261}{{\ttfamily 1902.07261}}].

\bibitem{Arvanitaki:2019rax}
A.~Arvanitaki, S.~Dimopoulos, M.~Galanis, L.~Lehner, J.~O. Thompson and
  K.~Van~Tilburg, \emph{{The Large-Misalignment Mechanism for the Formation of
  Compact Axion Structures: Signatures from the QCD Axion to Fuzzy Dark
  Matter}},  \href{https://arxiv.org/abs/1909.11665}{{\ttfamily 1909.11665}}.

\bibitem{Kawasaki:2020jnw}
M.~Kawasaki, W.~Nakano, H.~Nakatsuka and E.~Sonomoto, \emph{{Oscillons of
  Axion-Like Particle: Mass distribution and power spectrum}},
  \href{https://arxiv.org/abs/2010.09311}{{\ttfamily 2010.09311}}.

\bibitem{Dymnikova:2000dy}
I.~Dymnikova, L.~Koziel, M.~Khlopov and S.~Rubin, \emph{{Quasilumps from first
  order phase transitions}}, {\emph{Grav. Cosmol.} {\bfseries 6} (2000) 311}
  [\href{https://arxiv.org/abs/hep-th/0010120}{{\ttfamily hep-th/0010120}}].

\bibitem{Gleiser:2010qt}
M.~Gleiser, N.~Graham and N.~Stamatopoulos, \emph{{Long-Lived Time-Dependent
  Remnants During Cosmological Symmetry Breaking: From Inflation to the
  Electroweak Scale}},
  \href{https://doi.org/10.1103/PhysRevD.82.043517}{\emph{Phys. Rev. D}
  {\bfseries 82} (2010) 043517}
  [\href{https://arxiv.org/abs/1004.4658}{{\ttfamily 1004.4658}}].

\bibitem{Bond:2015zfa}
J.~R. Bond, J.~Braden and L.~Mersini-Houghton, \emph{{Cosmic bubble and domain
  wall instabilities III: The role of oscillons in three-dimensional bubble
  collisions}},
  \href{https://doi.org/10.1088/1475-7516/2015/09/004}{\emph{JCAP} {\bfseries
  09} (2015) 004} [\href{https://arxiv.org/abs/1505.02162}{{\ttfamily
  1505.02162}}].

\bibitem{Zhou:2013tsa}
S.-Y. Zhou, E.~J. Copeland, R.~Easther, H.~Finkel, Z.-G. Mou and P.~M. Saffin,
  \emph{{Gravitational Waves from Oscillon Preheating}},
  \href{https://doi.org/10.1007/JHEP10(2013)026}{\emph{JHEP} {\bfseries 10}
  (2013) 026} [\href{https://arxiv.org/abs/1304.6094}{{\ttfamily 1304.6094}}].

\bibitem{Helfer:2018vtq}
T.~Helfer, E.~A. Lim, M.~A. Garcia and M.~A. Amin, \emph{{Gravitational Wave
  Emission from Collisions of Compact Scalar Solitons}},
  \href{https://doi.org/10.1103/PhysRevD.99.044046}{\emph{Phys. Rev. D}
  {\bfseries 99} (2019) 044046}
  [\href{https://arxiv.org/abs/1802.06733}{{\ttfamily 1802.06733}}].

\bibitem{Liu:2017hua}
J.~Liu, Z.-K. Guo, R.-G. Cai and G.~Shiu, \emph{{Gravitational Waves from
  Oscillons with Cuspy Potentials}},
  \href{https://doi.org/10.1103/PhysRevLett.120.031301}{\emph{Phys. Rev. Lett.}
  {\bfseries 120} (2018) 031301}
  [\href{https://arxiv.org/abs/1707.09841}{{\ttfamily 1707.09841}}].

\bibitem{Amin:2018xfe}
M.~A. Amin, J.~Braden, E.~J. Copeland, J.~T. Giblin, C.~Solorio, Z.~J. Weiner
  et~al., \emph{{Gravitational waves from asymmetric oscillon dynamics?}},
  \href{https://doi.org/10.1103/PhysRevD.98.024040}{\emph{Phys. Rev. D}
  {\bfseries 98} (2018) 024040}
  [\href{https://arxiv.org/abs/1803.08047}{{\ttfamily 1803.08047}}].

\bibitem{Lozanov:2019ylm}
K.~D. Lozanov and M.~A. Amin, \emph{{Gravitational perturbations from oscillons
  and transients after inflation}},
  \href{https://doi.org/10.1103/PhysRevD.99.123504}{\emph{Phys. Rev.}
  {\bfseries D99} (2019) 123504}
  [\href{https://arxiv.org/abs/1902.06736}{{\ttfamily 1902.06736}}].

\bibitem{Dietrich:2018jov}
T.~Dietrich, F.~Day, K.~Clough, M.~Coughlin and J.~Niemeyer, \emph{{Neutron
  star\textendash{}axion star collisions in the light of multimessenger
  astronomy}}, \href{https://doi.org/10.1093/mnras/sty3158}{\emph{Mon. Not.
  Roy. Astron. Soc.} {\bfseries 483} (2019) 908}
  [\href{https://arxiv.org/abs/1808.04746}{{\ttfamily 1808.04746}}].

\bibitem{Hook:2018iia}
A.~Hook, Y.~Kahn, B.~R. Safdi and Z.~Sun, \emph{{Radio Signals from Axion Dark
  Matter Conversion in Neutron Star Magnetospheres}},
  \href{https://doi.org/10.1103/PhysRevLett.121.241102}{\emph{Phys. Rev. Lett.}
  {\bfseries 121} (2018) 241102}
  [\href{https://arxiv.org/abs/1804.03145}{{\ttfamily 1804.03145}}].

\bibitem{Clough:2018exo}
K.~Clough, T.~Dietrich and J.~C. Niemeyer, \emph{{Axion star collisions with
  black holes and neutron stars in full 3D numerical relativity}},
  \href{https://doi.org/10.1103/PhysRevD.98.083020}{\emph{Phys. Rev. D}
  {\bfseries 98} (2018) 083020}
  [\href{https://arxiv.org/abs/1808.04668}{{\ttfamily 1808.04668}}].

\bibitem{Levkov:2020txo}
D.~Levkov, A.~Panin and I.~Tkachev, \emph{{Radio-emission of axion stars}},
  \href{https://doi.org/10.1103/PhysRevD.102.023501}{\emph{Phys. Rev. D}
  {\bfseries 102} (2020) 023501}
  [\href{https://arxiv.org/abs/2004.05179}{{\ttfamily 2004.05179}}].

\bibitem{Prabhu:2020yif}
A.~Prabhu and N.~M. Rapidis, \emph{{Resonant Conversion of Dark Matter
  Oscillons in Pulsar Magnetospheres}},
  \href{https://arxiv.org/abs/2005.03700}{{\ttfamily 2005.03700}}.

\bibitem{Amin:2020vja}
M.~A. Amin and Z.-G. Mou, \emph{{Electromagnetic Bursts from Mergers of
  Oscillons in Axion-like Fields}},
  \href{https://arxiv.org/abs/2009.11337}{{\ttfamily 2009.11337}}.

\bibitem{Kawasaki:2020tbo}
M.~Kawasaki, W.~Nakano, H.~Nakatsuka and E.~Sonomoto, \emph{{Probing Oscillons
  of Ultra-Light Axion-like Particle by 21cm Forest}},
  \href{https://arxiv.org/abs/2010.13504}{{\ttfamily 2010.13504}}.

\bibitem{Helfer:2016ljl}
T.~Helfer, D.~J.~E. Marsh, K.~Clough, M.~Fairbairn, E.~A. Lim and R.~Becerril,
  \emph{{Black hole formation from axion stars}},
  \href{https://doi.org/10.1088/1475-7516/2017/03/055}{\emph{JCAP} {\bfseries
  03} (2017) 055} [\href{https://arxiv.org/abs/1609.04724}{{\ttfamily
  1609.04724}}].

\bibitem{Muia:2019coe}
F.~Muia, M.~Cicoli, K.~Clough, F.~Pedro, F.~Quevedo and G.~P. Vacca, \emph{{The
  Fate of Dense Scalar Stars}},
  \href{https://doi.org/10.1088/1475-7516/2019/07/044}{\emph{JCAP} {\bfseries
  07} (2019) 044} [\href{https://arxiv.org/abs/1906.09346}{{\ttfamily
  1906.09346}}].

\bibitem{Nazari:2020fmk}
Z.~Nazari, M.~Cicoli, K.~Clough and F.~Muia, \emph{{Oscillon collapse to black
  holes}},  \href{https://arxiv.org/abs/2010.05933}{{\ttfamily 2010.05933}}.

\bibitem{Widdicombe:2019woy}
J.~Y. Widdicombe, T.~Helfer and E.~A. Lim, \emph{{Black hole formation in
  relativistic Oscillaton collisions}},
  \href{https://doi.org/10.1088/1475-7516/2020/01/027}{\emph{JCAP} {\bfseries
  01} (2020) 027} [\href{https://arxiv.org/abs/1910.01950}{{\ttfamily
  1910.01950}}].

\bibitem{Khlopov:1985jw}
M.~Khlopov, B.~Malomed and I.~Zeldovich, \emph{{Gravitational instability of
  scalar fields and formation of primordial black holes}}, {\emph{Mon. Not.
  Roy. Astron. Soc.} {\bfseries 215} (1985) 575}.

\bibitem{Cotner:2018vug}
E.~Cotner, A.~Kusenko and V.~Takhistov, \emph{{Primordial Black Holes from
  Inflaton Fragmentation into Oscillons}},
  \href{https://doi.org/10.1103/PhysRevD.98.083513}{\emph{Phys. Rev.}
  {\bfseries D98} (2018) 083513}
  [\href{https://arxiv.org/abs/1801.03321}{{\ttfamily 1801.03321}}].

\bibitem{Cotner:2019ykd}
E.~Cotner, A.~Kusenko, M.~Sasaki and V.~Takhistov, \emph{{Analytic Description
  of Primordial Black Hole Formation from Scalar Field Fragmentation}},
  \href{https://doi.org/10.1088/1475-7516/2019/10/077}{\emph{JCAP} {\bfseries
  10} (2019) 077} [\href{https://arxiv.org/abs/1907.10613}{{\ttfamily
  1907.10613}}].

\bibitem{Segur:1987mg}
H.~Segur and M.~D. Kruskal, \emph{{Nonexistence of Small Amplitude Breather
  Solutions in $\phi^4$ Theory}},
  \href{https://doi.org/10.1103/PhysRevLett.58.747}{\emph{Phys. Rev. Lett.}
  {\bfseries 58} (1987) 747}.

\bibitem{Fodor:2008du}
G.~Fodor, P.~Forgacs, Z.~Horvath and M.~Mezei, \emph{{Computation of the
  radiation amplitude of oscillons}},
  \href{https://doi.org/10.1103/PhysRevD.79.065002}{\emph{Phys. Rev. D}
  {\bfseries 79} (2009) 065002}
  [\href{https://arxiv.org/abs/0812.1919}{{\ttfamily 0812.1919}}].

\bibitem{Hertzberg:2010yz}
M.~P. Hertzberg, \emph{{Quantum Radiation of Oscillons}},
  \href{https://doi.org/10.1103/PhysRevD.82.045022}{\emph{Phys. Rev.}
  {\bfseries D82} (2010) 045022}
  [\href{https://arxiv.org/abs/1003.3459}{{\ttfamily 1003.3459}}].

\bibitem{Mukaida:2016hwd}
K.~Mukaida, M.~Takimoto and M.~Yamada, \emph{{On Longevity of
  I-ball/Oscillon}}, \href{https://doi.org/10.1007/JHEP03(2017)122}{\emph{JHEP}
  {\bfseries 03} (2017) 122}
  [\href{https://arxiv.org/abs/1612.07750}{{\ttfamily 1612.07750}}].

\bibitem{Ibe:2019vyo}
M.~Ibe, M.~Kawasaki, W.~Nakano and E.~Sonomoto, \emph{{Decay of I-ball/Oscillon
  in Classical Field Theory}},
  \href{https://doi.org/10.1007/JHEP04(2019)030}{\emph{JHEP} {\bfseries 04}
  (2019) 030} [\href{https://arxiv.org/abs/1901.06130}{{\ttfamily
  1901.06130}}].

\bibitem{Zhang:2020bec}
H.-Y. Zhang, M.~A. Amin, E.~J. Copeland, P.~M. Saffin and K.~D. Lozanov,
  \emph{{Classical Decay Rates of Oscillons}},
  \href{https://doi.org/10.1088/1475-7516/2020/07/055}{\emph{JCAP} {\bfseries
  07} (2020) 055} [\href{https://arxiv.org/abs/2004.01202}{{\ttfamily
  2004.01202}}].

\bibitem{Grandclement:2011wz}
P.~Grandclement, G.~Fodor and P.~Forgacs, \emph{{Numerical simulation of
  oscillatons: extracting the radiating tail}},
  \href{https://doi.org/10.1103/PhysRevD.84.065037}{\emph{Phys. Rev. D}
  {\bfseries 84} (2011) 065037}
  [\href{https://arxiv.org/abs/1107.2791}{{\ttfamily 1107.2791}}].

\bibitem{Eby:2015hyx}
J.~Eby, P.~Suranyi and L.~Wijewardhana, \emph{{The Lifetime of Axion Stars}},
  \href{https://doi.org/10.1142/S0217732316500905}{\emph{Mod. Phys. Lett. A}
  {\bfseries 31} (2016) 1650090}
  [\href{https://arxiv.org/abs/1512.01709}{{\ttfamily 1512.01709}}].

\bibitem{Eby:2020ply}
J.~Eby, L.~Street, P.~Suranyi and L.~Wijewardhana, \emph{{Global View of Axion
  Stars with (Nearly) Planck-Scale Decay Constants}},
  \href{https://arxiv.org/abs/2011.09087}{{\ttfamily 2011.09087}}.

\bibitem{Seidel:1991zh}
E.~Seidel and W.~Suen, \emph{{Oscillating soliton stars}},
  \href{https://doi.org/10.1103/PhysRevLett.66.1659}{\emph{Phys. Rev. Lett.}
  {\bfseries 66} (1991) 1659}.

\bibitem{UrenaLopez:2002gx}
L.~Urena-Lopez, T.~Matos and R.~Becerril, \emph{{Inside oscillatons}},
  \href{https://doi.org/10.1088/0264-9381/19/23/320}{\emph{Class. Quant. Grav.}
  {\bfseries 19} (2002) 6259}.

\bibitem{Alcubierre:2003sx}
M.~Alcubierre, R.~Becerril, S.~F. Guzman, T.~Matos, D.~Nunez and L.~A.
  Urena-Lopez, \emph{{Numerical studies of Phi**2 oscillatons}},
  \href{https://doi.org/10.1088/0264-9381/20/13/332}{\emph{Class. Quant. Grav.}
  {\bfseries 20} (2003) 2883}
  [\href{https://arxiv.org/abs/gr-qc/0301105}{{\ttfamily gr-qc/0301105}}].

\bibitem{Visinelli:2017ooc}
L.~Visinelli, S.~Baum, J.~Redondo, K.~Freese and F.~Wilczek, \emph{{Dilute and
  dense axion stars}},
  \href{https://doi.org/10.1016/j.physletb.2017.12.010}{\emph{Phys. Lett. B}
  {\bfseries 777} (2018) 64}
  [\href{https://arxiv.org/abs/1710.08910}{{\ttfamily 1710.08910}}].

\bibitem{Eby:2019ntd}
J.~Eby, M.~Leembruggen, L.~Street, P.~Suranyi and L.~R. Wijewardhana,
  \emph{{Global view of QCD axion stars}},
  \href{https://doi.org/10.1103/PhysRevD.100.063002}{\emph{Phys. Rev. D}
  {\bfseries 100} (2019) 063002}
  [\href{https://arxiv.org/abs/1905.00981}{{\ttfamily 1905.00981}}].

\bibitem{Hui:2016ltb}
L.~Hui, J.~P. Ostriker, S.~Tremaine and E.~Witten, \emph{{Ultralight scalars as
  cosmological dark matter}},
  \href{https://doi.org/10.1103/PhysRevD.95.043541}{\emph{Phys. Rev. D}
  {\bfseries 95} (2017) 043541}
  [\href{https://arxiv.org/abs/1610.08297}{{\ttfamily 1610.08297}}].

\bibitem{Kallosh:2013hoa}
R.~Kallosh and A.~Linde, \emph{{Universality Class in Conformal Inflation}},
  \href{https://doi.org/10.1088/1475-7516/2013/07/002}{\emph{JCAP} {\bfseries
  1307} (2013) 002} [\href{https://arxiv.org/abs/1306.5220}{{\ttfamily
  1306.5220}}].

\bibitem{Silverstein:2008sg}
E.~Silverstein and A.~Westphal, \emph{{Monodromy in the CMB: Gravity Waves and
  String Inflation}},
  \href{https://doi.org/10.1103/PhysRevD.78.106003}{\emph{Phys. Rev.}
  {\bfseries D78} (2008) 106003}
  [\href{https://arxiv.org/abs/0803.3085}{{\ttfamily 0803.3085}}].

\bibitem{McAllister:2014mpa}
L.~McAllister, E.~Silverstein, A.~Westphal and T.~Wrase, \emph{{The Powers of
  Monodromy}}, \href{https://doi.org/10.1007/JHEP09(2014)123}{\emph{JHEP}
  {\bfseries 09} (2014) 123} [\href{https://arxiv.org/abs/1405.3652}{{\ttfamily
  1405.3652}}].

\bibitem{Guth:2014hsa}
A.~H. Guth, M.~P. Hertzberg and C.~Prescod-Weinstein, \emph{{Do Dark Matter
  Axions Form a Condensate with Long-Range Correlation?}},
  \href{https://doi.org/10.1103/PhysRevD.92.103513}{\emph{Phys. Rev. D}
  {\bfseries 92} (2015) 103513}
  [\href{https://arxiv.org/abs/1412.5930}{{\ttfamily 1412.5930}}].

\bibitem{Hindmarsh:2006ur}
M.~Hindmarsh and P.~Salmi, \emph{{Numerical investigations of oscillons in 2
  dimensions}}, \href{https://doi.org/10.1103/PhysRevD.74.105005}{\emph{Phys.
  Rev. D} {\bfseries 74} (2006) 105005}
  [\href{https://arxiv.org/abs/hep-th/0606016}{{\ttfamily hep-th/0606016}}].

\bibitem{Sanchis-Gual:2019ljs}
N.~Sanchis-Gual, F.~Di~Giovanni, M.~Zilhão, C.~Herdeiro, P.~Cerdá-Durán,
  J.~Font et~al., \emph{{Nonlinear Dynamics of Spinning Bosonic Stars:
  Formation and Stability}},
  \href{https://doi.org/10.1103/PhysRevLett.123.221101}{\emph{Phys. Rev. Lett.}
  {\bfseries 123} (2019) 221101}
  [\href{https://arxiv.org/abs/1907.12565}{{\ttfamily 1907.12565}}].

\bibitem{Friedberg:1976me}
R.~Friedberg, T.~Lee and A.~Sirlin, \emph{{A Class of Scalar-Field Soliton
  Solutions in Three Space Dimensions}},
  \href{https://doi.org/10.1103/PhysRevD.13.2739}{\emph{Phys. Rev. D}
  {\bfseries 13} (1976) 2739}.

\bibitem{Lee:1991ax}
T.~Lee and Y.~Pang, \emph{{Nontopological solitons}},
  \href{https://doi.org/10.1016/0370-1573(92)90064-7}{\emph{Phys. Rept.}
  {\bfseries 221} (1992) 251}.

\bibitem{Gibbons:1976ue}
G.~Gibbons and S.~Hawking, \emph{{Action Integrals and Partition Functions in
  Quantum Gravity}},
  \href{https://doi.org/10.1103/PhysRevD.15.2752}{\emph{Phys. Rev. D}
  {\bfseries 15} (1977) 2752}.

\bibitem{Lee:1988av}
T.~D. Lee and Y.~Pang, \emph{{Stability of Mini-Boson Stars}},
  \href{https://doi.org/10.1016/0550-3213(89)90365-9}{\emph{Nucl. Phys. B}
  {\bfseries 315} (1989) 477}.

\bibitem{Arnowitt:1962hi}
R.~L. Arnowitt, S.~Deser and C.~W. Misner, \emph{{The Dynamics of general
  relativity}}, \href{https://doi.org/10.1007/s10714-008-0661-1}{\emph{Gen.
  Rel. Grav.} {\bfseries 40} (2008) 1997}
  [\href{https://arxiv.org/abs/gr-qc/0405109}{{\ttfamily gr-qc/0405109}}].

\bibitem{Eby:2018ufi}
J.~Eby, K.~Mukaida, M.~Takimoto, L.~C.~R. Wijewardhana and M.~Yamada,
  \emph{{Classical nonrelativistic effective field theory and the role of
  gravitational interactions}},
  \href{https://doi.org/10.1103/PhysRevD.99.123503}{\emph{Phys. Rev.}
  {\bfseries D99} (2019) 123503}
  [\href{https://arxiv.org/abs/1807.09795}{{\ttfamily 1807.09795}}].

\bibitem{Braaten:2018lmj}
E.~Braaten, A.~Mohapatra and H.~Zhang, \emph{{Classical Nonrelativistic
  Effective Field Theories for a Real Scalar Field}},
  \href{https://doi.org/10.1103/PhysRevD.98.096012}{\emph{Phys. Rev. D}
  {\bfseries 98} (2018) 096012}
  [\href{https://arxiv.org/abs/1806.01898}{{\ttfamily 1806.01898}}].

\bibitem{Namjoo:2017nia}
M.~H. Namjoo, A.~H. Guth and D.~I. Kaiser, \emph{{Relativistic Corrections to
  Nonrelativistic Effective Field Theories}},
  \href{https://doi.org/10.1103/PhysRevD.98.016011}{\emph{Phys. Rev. D}
  {\bfseries 98} (2018) 016011}
  [\href{https://arxiv.org/abs/1712.00445}{{\ttfamily 1712.00445}}].

\bibitem{Salehian:2020bon}
B.~Salehian, M.~H. Namjoo and D.~I. Kaiser, \emph{{Effective theories for a
  nonrelativistic field in an expanding universe: Induced self-interaction,
  pressure, sound speed, and viscosity}},
  \href{https://arxiv.org/abs/2005.05388}{{\ttfamily 2005.05388}}.

\bibitem{Friedberg:1986tp}
R.~Friedberg, T.~Lee and Y.~Pang, \emph{{MINI - SOLITON STARS}},
  \href{https://doi.org/10.1103/PhysRevD.35.3640}{\emph{Phys. Rev. D}
  {\bfseries 35} (1987) 3640}.

\bibitem{Alcubierre:2000xu}
M.~Alcubierre, G.~Allen, B.~Bruegmann, T.~Dramlitsch, J.~A. Font,
  P.~Papadopoulos et~al., \emph{{Towards a stable numerical evolution of
  strongly gravitating systems in general relativity: The Conformal
  treatments}}, \href{https://doi.org/10.1103/PhysRevD.62.044034}{\emph{Phys.
  Rev.} {\bfseries D62} (2000) 044034}
  [\href{https://arxiv.org/abs/gr-qc/0003071}{{\ttfamily gr-qc/0003071}}].

\bibitem{Zlochower:2005bj}
Y.~Zlochower, J.~Baker, M.~Campanelli and C.~Lousto, \emph{{Accurate black hole
  evolutions by fourth-order numerical relativity}},
  \href{https://doi.org/10.1103/PhysRevD.72.024021}{\emph{Phys. Rev. D}
  {\bfseries 72} (2005) 024021}
  [\href{https://arxiv.org/abs/gr-qc/0505055}{{\ttfamily gr-qc/0505055}}].

\bibitem{alcubierre2008introduction}
M.~Alcubierre, \emph{Introduction to 3+ 1 numerical relativity}, vol.~140.
  Oxford University Press, 2008.

\end{thebibliography}\endgroup
\end{document}